\documentclass[prb,twocolumn,showpacs,amsmath,amssymb]{revtex4}

\usepackage{bm}
\usepackage[dvipdfmx]{graphicx}
\usepackage{color}

\begin{document}

\title{Manipulating quantum channels in weak topological insulator nanoarchitectures}

\author{Akihiko Matsumoto}
\author{Takashi Arita}
\author{Yositake Takane}
\author{Yukinori Yoshimura}
\author{Ken-Ichiro Imura}

\affiliation{$^1$Department of Quantum Matter, AdSM, Hiroshima University, Higashi-Hiroshima, 739-8530, Japan}

\date{\today}

\begin{abstract}
In {\it strong} topological insulators
protected surface states are always manifest,
while in {\it weak} topological insulators (WTI) 
the corresponding
metallic surface states are 
either manifest or hidden, depending on the orientation of the surface.
One can design a nanostep on the surface of WTI
such that a protected helical channel appears 
along it.
In a more generic WTI nanostructure,
multiple sets of such quasi-1D channels
emerge and are coupled to each other.
We study the response of
the electronic spectrum associated with such quasi-1D surface modes
against a magnetic flux piercing the system 
in the presence of disorder,
and find a non-trivial, connected spectral flow as
a clear signature indicating the immunity of the surface modes to disorder.
We propose
that the WTI nanoarchitecture is a promising platform for realizing
topologically protected nanocircuits immune to disorder.
\end{abstract}


\maketitle

\section{Introduction}

Three-dimensional (3D) topological insulators
are classified into weak and strong.
\cite{MooreBalents, FuKaneMele, Roy}
The strong topological insulator (STI) 
exhibits a single Dirac cone in the surface Brillouin zone (BZ),
which is immune to backscattering by non-magnetic impurities.
\cite{KN, Bardarson}
The immunity to backscattering also implies that
an electronic state in the single Dirac cone
cannot be confined in a finite area.
Instead, it is extended to the entire surface of STI,\cite{kado}
making all its facets metallic.
On contrary,
the weak topological insulator (WTI) 
exhibits an even number of (typically two)
Dirac cones in the surface BZ,
which can be confined;
among surfaces of a WTI sample oriented in different directions
there are also {\it gapped} surfaces.
In this sense
topological non-triviality is always manifest in STI, 
while in WTI it is either manifest or hidden.
\cite{Ran_nphys, disloc, strong_side, Liu_physica, mayuko1, dark, TKN14, TI_nanofilms}
Besides, in WTI
one can actually switch it on and off.
We have previously shown \cite{dark}
that a one-dimensional (1D) helical channel emerges along a step formed 
on the surface of a WTI,
and it can be regarded as a perfectly conducting channel (PCC)
without backscattering.
Using such PCCs,
one can possibly construct a nano-circuit
of protected 1D helical modes
on the surface of WTI 
by simply patterning it with the use of lithography and etching.
This controllability of the topological non-trivialness 
sometimes makes WTI more {\it useful} than STI.

An experimental realization of a WTI has been reported in a bismuth-based
layered compound Bi$_{14}$Rh$_3$I$_9$. \cite{WTI_exp}
More recently,
a helical 1D channel emergent on a step-like surfaces of a WTI \cite{dark}
has been also observed experimentally.\cite{Morgen}
Yet, in spite of the number of realizations of
3D topological insulators,
\cite{Ando}
there have not been many proposals for realizing a WTI
in stoichiometric compounds.
\cite{BiTeI}
WTI may be realized in superlattice systems.
\cite{Claudia, FIH, TSI_theo, WCI}
Features specific to WTI can be also seen in the so-called
topological crystalline insulators (TCI).
\cite{Fu_TCI, TCI_Nature2, TCI_Nature1, TCI_Nature3}
Unlike the standard topological insulators protected by time-reversal symmetry,
TCI is protected by crystalline symmetry.\cite{Morimoto}

In Ref. 12, the case of a single nano-circuit emergent on a WTI surface 
has been analyzed in some detail. 
In reality, however, 
in the case of any realistic nano-circuit useful for application,
there would be more than a single circuit, interacting with each other on the chip. 
Here, in this paper we highlight, in contrast to Ref. \onlinecite{dark}, 
such multi-channel cases.
As a typical setup for realizing a mentioned PCC
we consider patterned surfaces of a WTI film or a flake;
panel (a) of Fig. 1 represents an idealized example of such nano-flake. 
We need a step of height corresponding to 
an odd number of atomic layers 
formed on a gapped surface
[see panel (b) of Fig. 1];
a robust PCC is guaranteed to exist in this case
(see Sec. II-A for a more detailed description).
If the height of the step is even, the channel tends to get gapped and localized. 
Such an even/odd feature has been also studied in Ref. \onlinecite{dark}, 
and the arguments given there
can be used to characterize
a single isolated channel. 
However, 
a circuit of such a PCC is inevitably closed on a surface of WTI nanoflake 
reflecting its topological nature.
Then, one has to consider not only the height of a step 
but also that of the base nano-flake. 
As an idealistic example we first consider the case of a single step 
on such a nano-flake (see Sec. IV A), 
before considering the more interesting case of two steps as a nontrivial example 
(Sec. IV B).

So far
we have in mind
the cases of single and double PCC nano-circuits,
discussed respectively in Sec. IV A and in Sec. IV B.
We analyze in these sections how they respond to disorder,
localized vs. delocalized, etc. 
By studying response of the system 
against flux insertion numerically,
we characterize quantum junctions
at which multiple sets of quasi-1D channels meet and couple to one another
under a certain ``traffic rule''.
Nano-circuits formed on the surface of the WTI sample
are composed of different parts
incident either at the step
or on side surfaces of the sample.
They meet typically at either end of the step region,
forming a junction of quantum channels.
At such junctions
they are connected under the rule revealed
by the above flux-insertion numerical experiment:
some are strongly (smoothly) connected
to form a part of PCC, others not.
As a concrete example of double PCC we 
consider
a rather specific geometry with two steps;
one  on the top and the other at the bottom of a nano-flake, 
but the obtained traffic rules could be equally applied to 
a more generic but topologically equivalent geometries.
In a generic situation we will have multiple sets of such
nano-circuits.
Yet, based on the observation we establish here
in the single and double PCC cases,
we can naturally conjecture that characteristics of such multiple setup 
can be reduced to those of the single and double PCC nano-circuits. 
In the same sense that the single and double PCC nano-circuits are 
immune to backscattering by disorder, 
a more generic setup with multiple PCCs is also considered 
to be immune to disorder.

The paper is organized as follows. 
In Sec. II we introduce
and define our model
employed in the (tight-binding) numerical simulation
presented in the subsequent sections.
In Sec. III
we discuss even/odd features in the case of 
a WTI nano-flake.
In Sec. IV,
we extend this observation to characterize
different variations of the 1D helical modes
emergent in WTI nano-structures,
typically at a step or at steps.
It is shown to be possible to 
realize various types of
non-trivial junctions that involve 
1D PCCs
in WTI nano-structures.
Sec. V is devoted to conclusions.

\begin{figure}[htbp]
\includegraphics[width=80mm, bb=0 0 857 733]{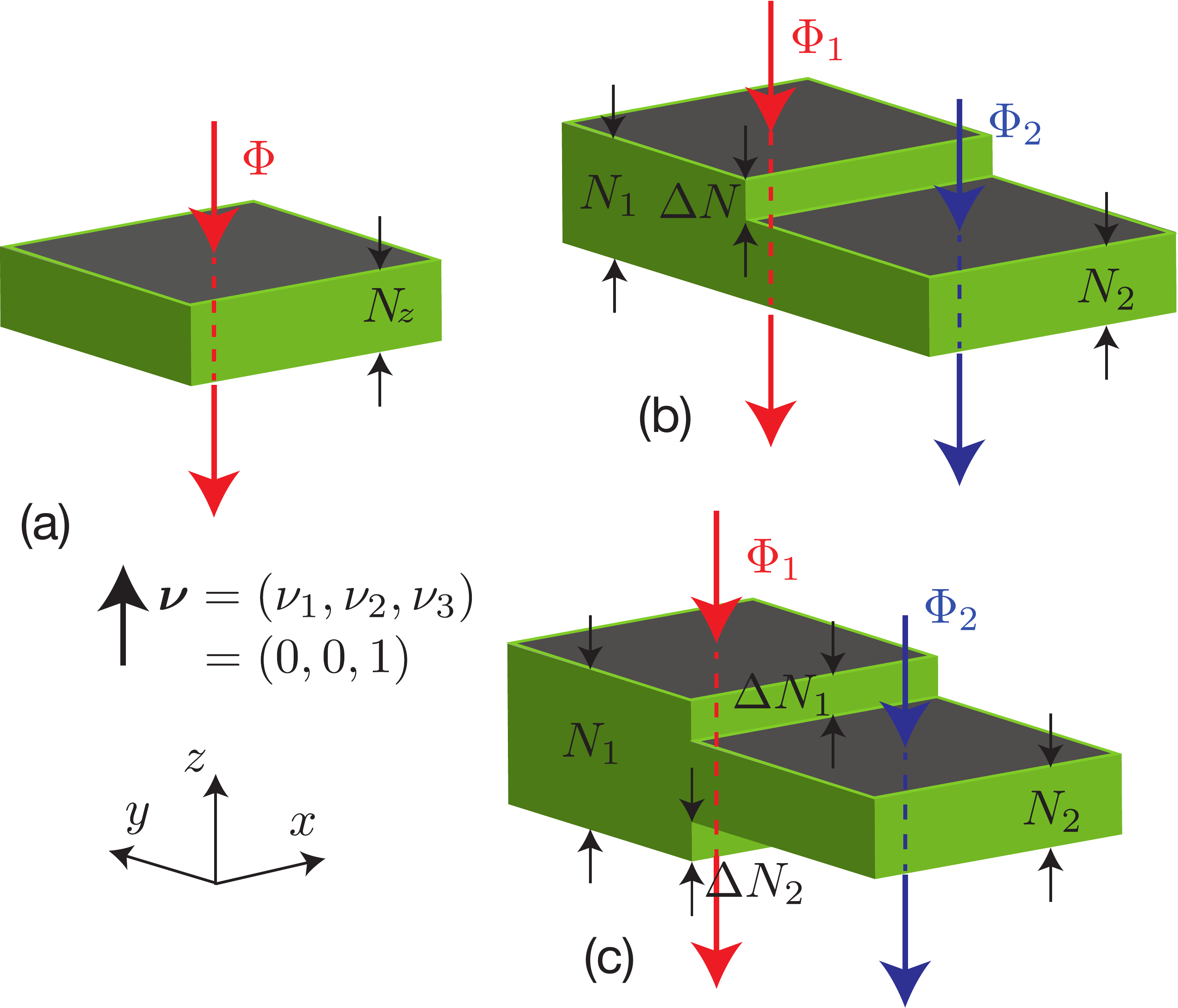}
\caption{Simplest examples of the weak topological insulator nano-architecture.
(a) 
A rectangular nano-flake 
geometry; open boundary conditions in all the three ($x$-, $y$- and $z$-) directions.
The top and the bottom surfaces are {\it gapped}.
(b) A step or (c) steps formed at the junction of two prisms of different heights
(see Sec. IV-B for details).
To quantify the robustness of the surface states that emerge in such
nanostructures, we study response of the system against
a magnetic flux inserted as shown in the figure 
[$\Phi$ in panel (a), or $\Phi_1$ and $\Phi_2$ in panels (b), (c)].}
\label{schema}
\end{figure}

\section{Model}

To study the robustness of
1D helical channels emergent on the surface of a WTI nano-structure,
we first need to {\it design} such a nano-structure.
We define our bulk effective Hamiltonian,
introduce the type of random potentials distributed over the sample,
and then sketch our standard recipe for performing a numerical simulation.

\subsection{Bulk material: a specific type of time-reversal invariant $\mathbb{Z}_2$ topological insulator}

Our bulk topological insulator (TI) is a standard type 
protected by the time reversal symmetry (TRS).
It is also called a $\mathbb{Z}_2$TI,
since it is distinguished from the trivial band insulator by 
a $\mathbb{Z}_2$ type topological number. \cite{KaneMele_Z2}
Compared with the trivial band insulator it has an inverted band gap;
this inversion is usually due to a strong spin-orbit coupling which preserves TRS.
When the primary $\mathbb{Z}_2$ index $\nu_0$ is nonzero ({i.e.}, $=1$),
such a TI is called a ``strong'' TI and exhibits a single Dirac cone on all the facets of the sample.
When $\nu_0=0$,
we still have a possibility that the insulator is nontrivial;
the secondary $\mathbb{Z}_2$ indices $\nu_1$, $\nu_2$, $\nu_3$
could be still nonzero.
In the reminder of the paper we focus on cases in which our bulk material falls on the class of
such a ``weak'' TI or WTI.
The WTI exhibits generally an even number of Dirac cones on its surface,
while the surface normal to the direction $\bm\nu = (\nu_1, \nu_2, \nu_3)$
exhibits no gapless Dirac cone.

\subsection{Model geometries}

As a toy example of such a nano-architecture,
we first consider
the case of a WTI nano-flake, 
as represented in Fig. \ref{schema} (a).
In the figure, the film thickness is somewhat exaggerated for clarity,
and here
we consider a typical and realistic situation
\cite{WTI_exp}
in which the top and bottom surfaces of the film
are {\it gapped}.
As shown in the figure,
this has been encoded in the choice of weak indices:
$\bm\nu =(\nu_1, \nu_2, \nu_3)= (0,0,1)$.
For this case, the WTI nanofilm consists
of unit atomic layers stacked in the $z$-direction, 
where each layer can be regarded as 
a 2D quantum spin-Hall (QSH) insulator 
having a counter-propagating pair of 1D (gapless) helical edge channels. 
Let $N_z$ be the number of unit atomic layers.
In the limit of decoupled such 2D QSH states there are $N_z$ 
pairs of 1D helical edge channels,
each circulating around the corresponding 2D QSH layer.
When such $N_z$ 2D QSH layers are coupled to form a WTI in the bulk,
the corresponding 1D channels
are also coupled to form surface states of the bulk WTI
which appears only on side surfaces of the film.
In the film geometry,
here represented as a flattened rectangular prism of height $N_z$,
the conducting property of such WTI surface states
is a drastic function of $N_z$, since 
they are confined in a finite width $N_z$
of side surfaces.
\cite{takane, strong_side, mayuko1,TI_nanofilms}
If $N_z$ is odd, there always remain, after recombination,
a single pair of helical modes
that are gapless, extended
and perfectly conducting:
formation of the PCC in the case of $N_z$ odd.
If $N_z$ is even,
all the channels are gapped and tend to get localized.

Once the electronic properties of such WTI films are properly addressed,
we proceed to analyzing 
the case of WTI terraces.
In Fig. \ref{schema} (b)
a simplest example of such terraces is
modeled by
two prisms of different heights $N_1$ and $N_2$
joined together through a side surface.
A scenario similar to the nanofilm case applies 
to this case of single step geometry;
when the height of the step is an odd-integer multiple of atomic layers,
there appears a robust 1D channel along the step.
\cite{dark}
If one thinks of a more generic WTI nanostructure,
multiple sets of such 1D channels are expected to appear,
and couple to each other.
If isolated, each set of 1D channels acts according to the even/odd rule
we find in the case of the prism,
while when they get together, interact, and eventually
recombine,
it is less trivial to tell what would happen.

In panel (c) of Fig. \ref{schema}
we give an example of multiple
1D channels along steps that run in parallel; one stemming from the top, the other from the bottom
surface. 
Let us assume that
$\Delta N_1$ and $\Delta N_2$ are both odd,
giving rise to 1D channels that are robust against disorder.
If $N_1$ and $N_2$ are also odd,
the robust 1D channels are extended to side surfaces either 
on the $N_1$ or to the $N_2$ side.
An interesting question is how
the two channels incident at the step
recombine to either of the side surfaces 
at the junction of quantum channels 
formed at both ends of the step.
If two channels at the step
are spatially well separated ($N_2 \gg 1$)
they do not interact and will act as two independent, perfectly conducting channel.
In the opposite limit: $N_2\sim 1$,
we address in this paper,
two odd-number channels run in parallel close to each other.
In this situation it is {\it a priori} not clear whether or not 
this pair of an {\it odd}-number of channels
merge together to become a single set of gapped 
{\it even}-number channels.
To probe the nature of helical modes
along the step and around the side surfaces,
we study response of the system to a magnetic flux.
We try different ways of inserting a flux:
{\rm e.g.}, 
$\Phi_1$ and $\Phi_2$
in Fig. \ref{schema} (c)
to obtain further insight on
how different  1D channels couple
and recombine to each another.

\subsection{Effective Hamiltonian, model parameters}

To represent
a bulk TI we consider the following Wilson-Dirac type
effective Hamiltonian
\cite{Liu_nphys, Liu_PRB}
\begin{equation}
h(\bm k)  = 
\tau_z m(\bm k) + \tau_x \sigma_\mu A_\mu \sin k_\mu,
\label{h_sink}
\end{equation}
where
\begin{equation}
m(\bm k) = m_0 + 2 m_{2\mu} (1-\cos k_\mu).
\label{m_cosk}
\end{equation}
In Eqs. (\ref{h_sink}) and (\ref{m_cosk})
a summation over the repeated index $\mu=x,y,z$
is not shown explicitly.
Eq. (\ref{h_sink}) can be regarded as a $4\times 4$ matrix,
spanned by two types of Pauli matrices $\bm \sigma$ and $\bm \tau$
each representing physically real and orbital spins.
Eq. (\ref{h_sink})
can be regarded as
a lattice version of
the continuum Dirac Hamiltonian
\begin{equation}
h_\Gamma (\bm k) = \tau_z m(\bm k) + \tau_x \sigma_\mu A_\mu k_\mu,
\label{h_k}
\end{equation}
at the $\Gamma$-point,
where
$m(\bm k) = m_0 + m_{2\mu} k_\mu^2$.
Here, in Eqs. (\ref{h_sink}), (\ref{m_cosk})
the lattice is chosen to be simple cubic for simplicity.
On top of Eqs. (\ref{h_sink}), (\ref{m_cosk})
we also consider on-site potential disorder of strength $W$.
On each site $(x,y,z)$ of the cubic lattice
a random potential of magnitude $V(x,y,z)$ is introduced, 
and distributed uniformly in the range of $[-W/2, W/2]$.


By varying the mass parameters in Eq. (\ref{m_cosk}) 
one can realize various weak and strong TI phases
\cite{mayuko1}
characterized by
strong and weak indices, $\nu_0$ and 
$\bm\nu=(\nu_1, \nu_2, \nu_3)$.
Here, to achieve a situation in which
the weak vector $\bm\nu$ is given by $\bm\nu =(0,0,1)$,
we choose the mass parameters such that
\begin{eqnarray}
m_{2x} =  m_{2y}=m_{2\parallel},\ \ \ 
m_{2z}/m_{2\parallel}= 0.1,
\nonumber \\
m_0/m_{2\parallel} = - 2.
\label{param}
\end{eqnarray}

\subsection{Numerical simulations}

The tight-binding form of our model Hamiltonian
 (\ref{h_sink})
is also useful for implementing the real space geometries
introduced in Sec. A.
Indeed, 
Eq. (\ref{h_sink}) represents
a tight-binding Hamiltonian 
with only onsite and nearest neighbor hopping terms 
defined on the cubic lattice.
Then,
the real space geometries as depicted in the three panels of Fig. \ref{schema}
can be implemented by
setting all the hopping parameters outside the designed nano-structure to be null
in the real space representation of 
Eq. (\ref{h_sink}).
In the actual implementation
we can simply truncate the real space Hamiltonian into a $4N_s\times 4N_s$ matrix,
where $N_s$ is the number of sites 
representing the nano-structure.

\begin{figure*}[htbp]
(a)
\includegraphics[width=70mm, bb=0 0 260 260]{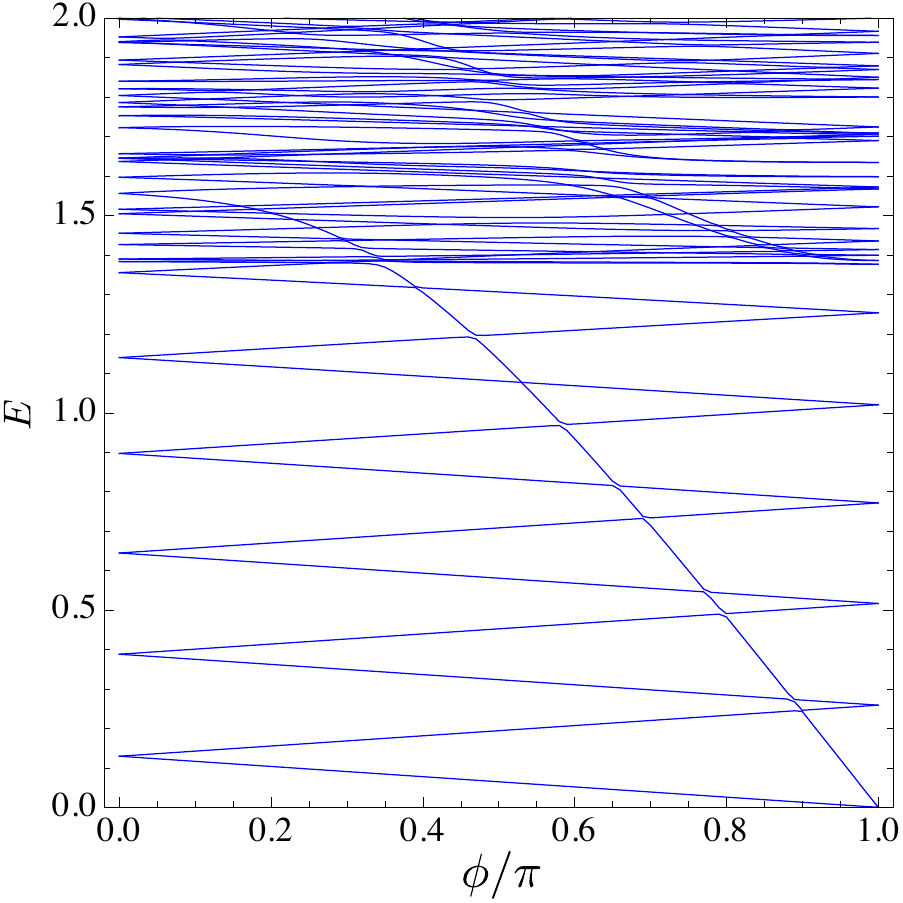}
(b)
\includegraphics[width=70mm, bb=0 0 260 260]{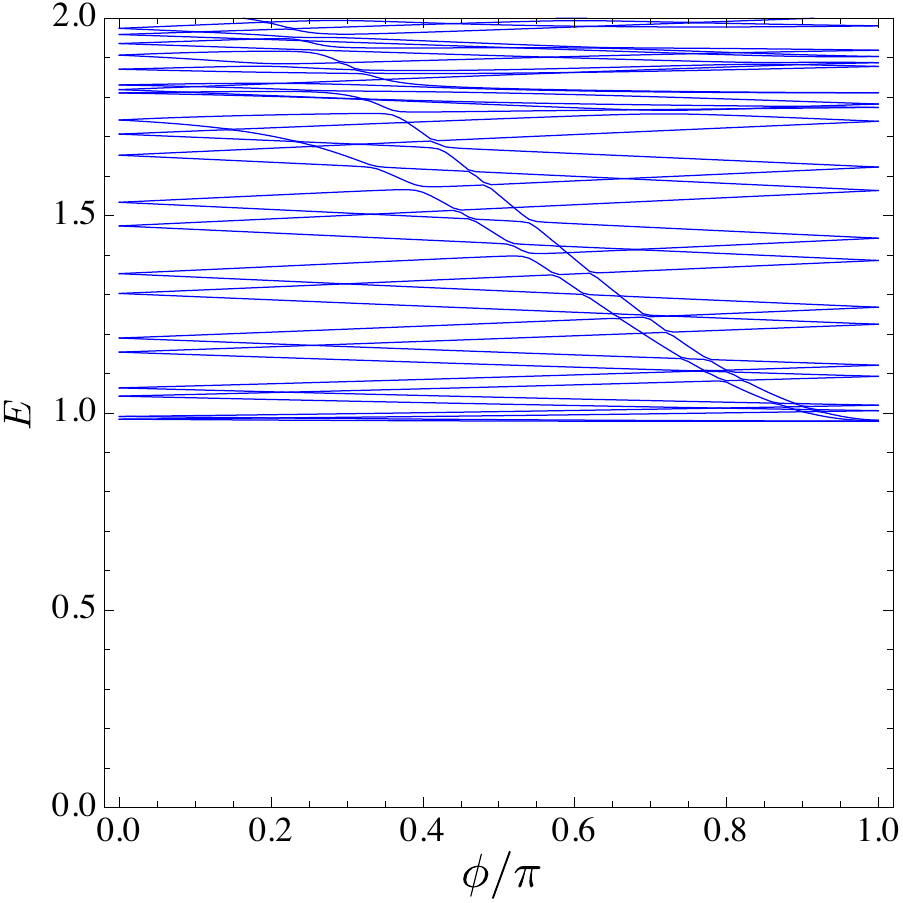}
\\
(c)
\includegraphics[width=70mm, bb=0 0 260 260]{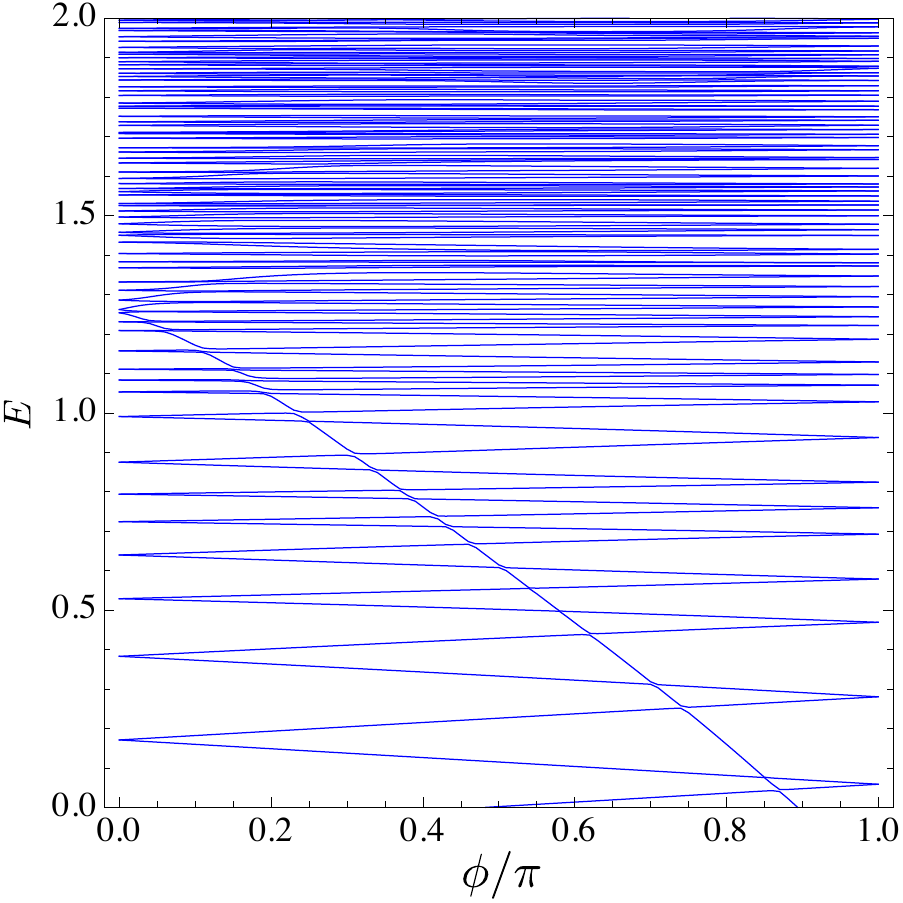}
(d)
\includegraphics[width=70mm, bb=0 0 260 260]{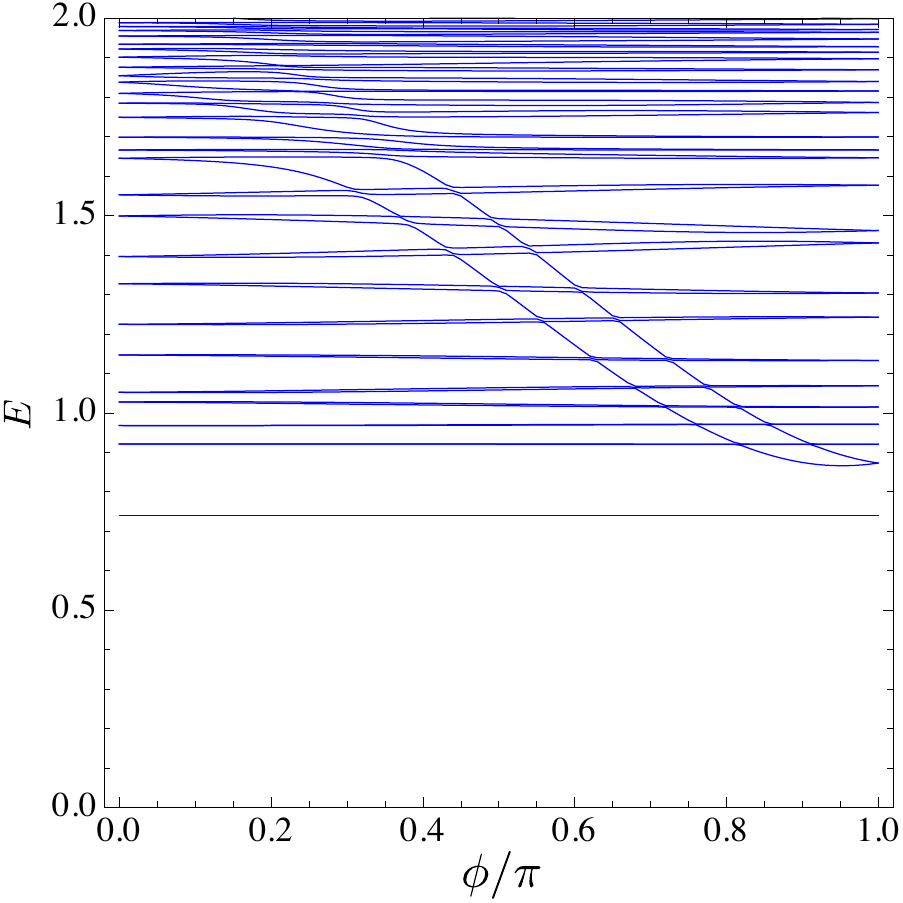}
\vspace{-2mm}
\caption{Spectral flow in WTI nanofilms.
Evolution of the spectrum 
$E(\phi)$ 
as a function of the flux $\phi = 2\pi (\Phi/\Phi_0)$
is shown in the clean limit ($W=0$) [panels (a), (b)],
and at moderate disorder ($W=2$) [panels (c), (d)].
Panels (a), (c) are for the case of $N_z$ odd ($N_z=3$),
while (b), (d) are for the case of $N_z$ even ($N_z=2$).
}
\label{film}
\end{figure*}

\begin{figure*}[htbp]
(a)
\includegraphics[width=70mm, bb =0 0 426 245]{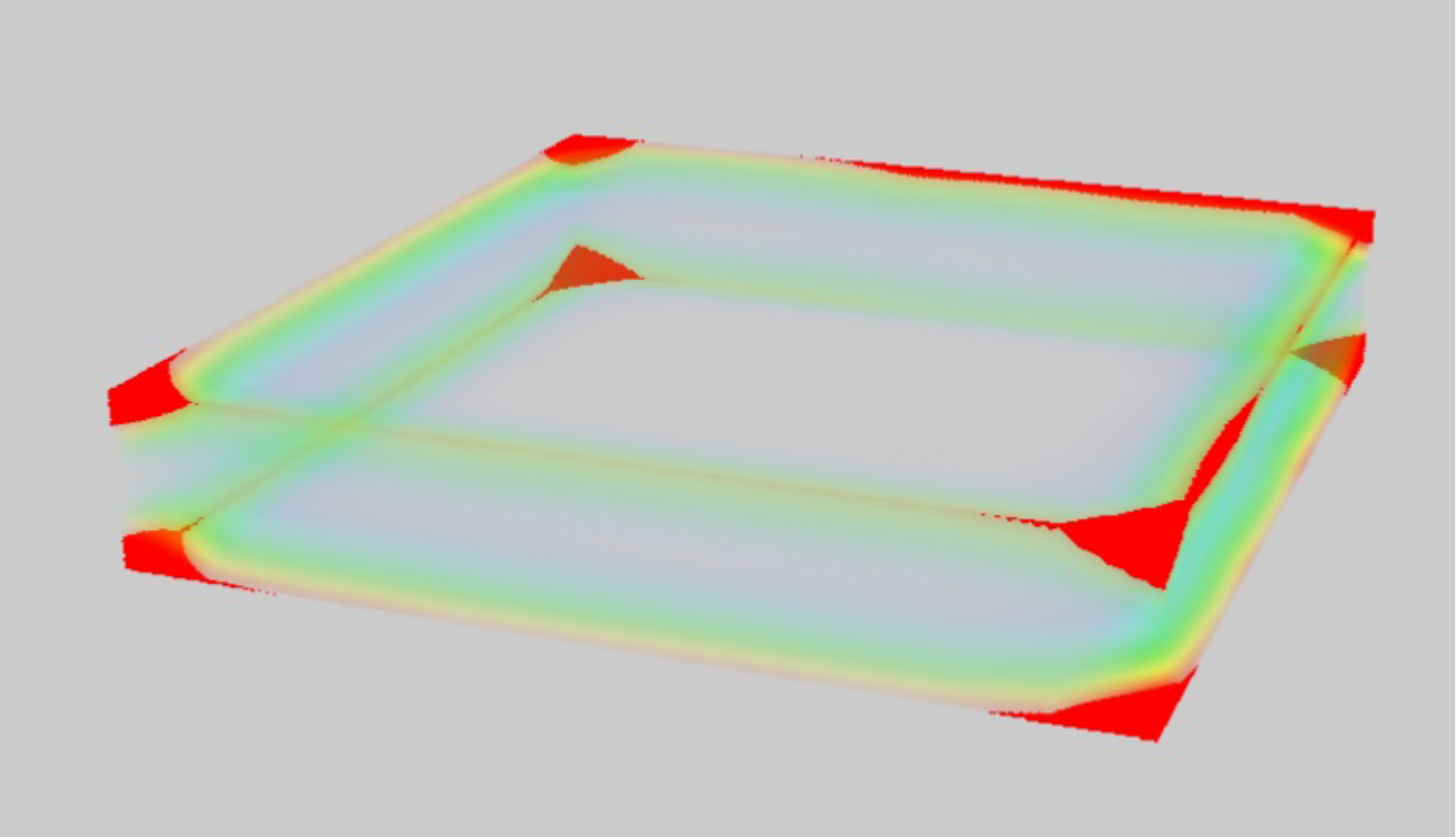}
(b)
\includegraphics[width=70mm, bb =0 0 426 229]{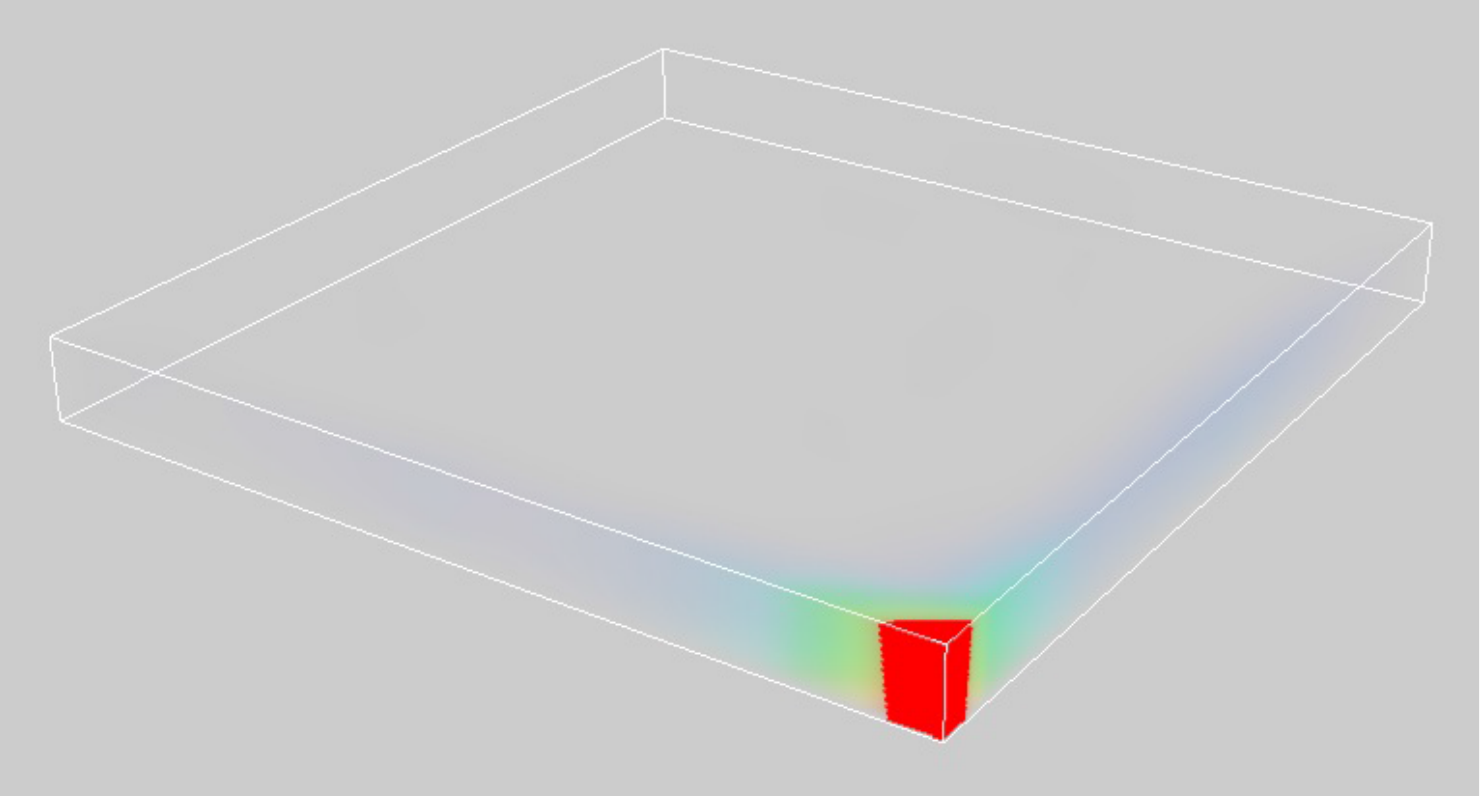}
\\
(c)
\includegraphics[width=70mm, bb =0 0 404 236]{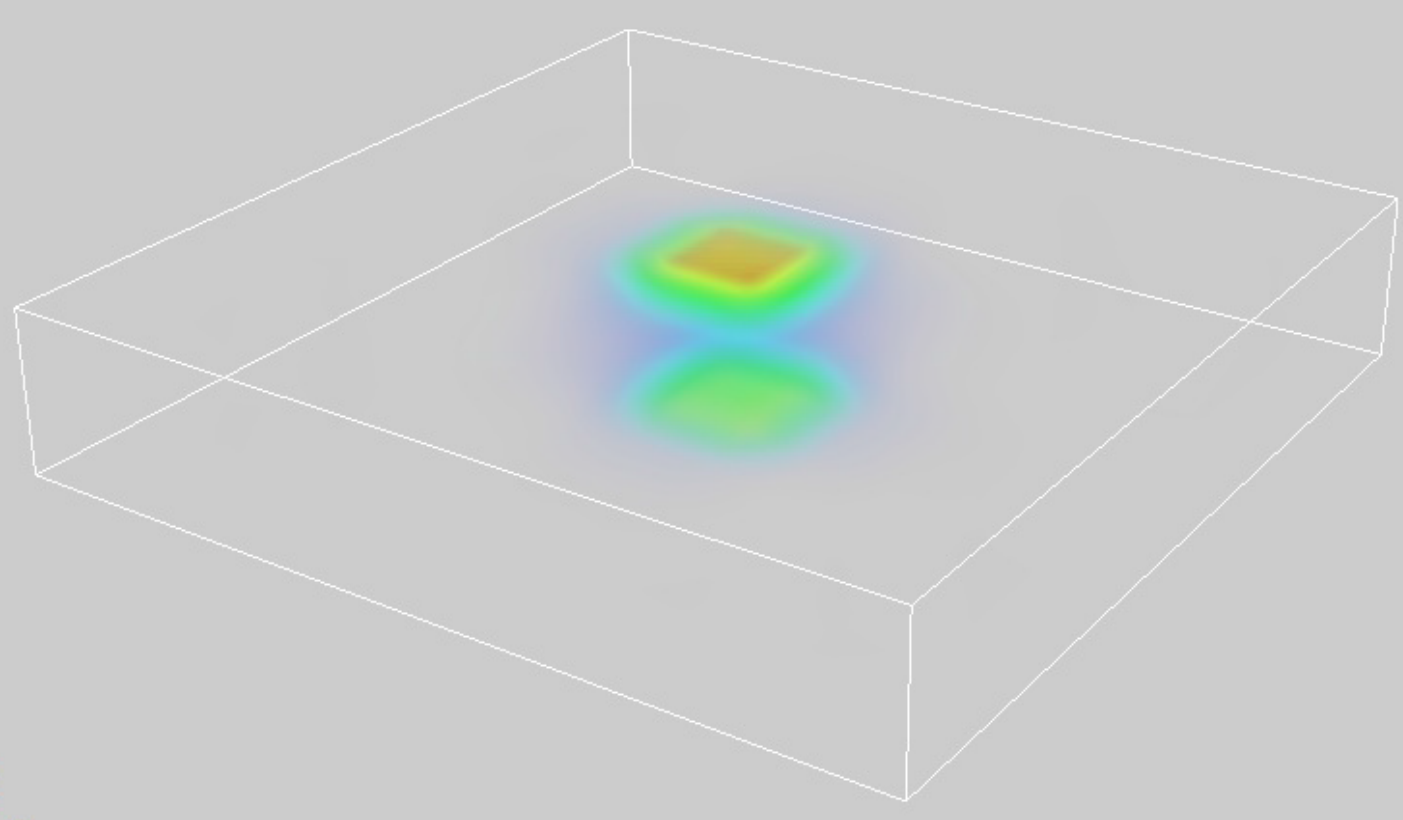}
(d)
\includegraphics[width=70mm, bb =0 0 411 216]{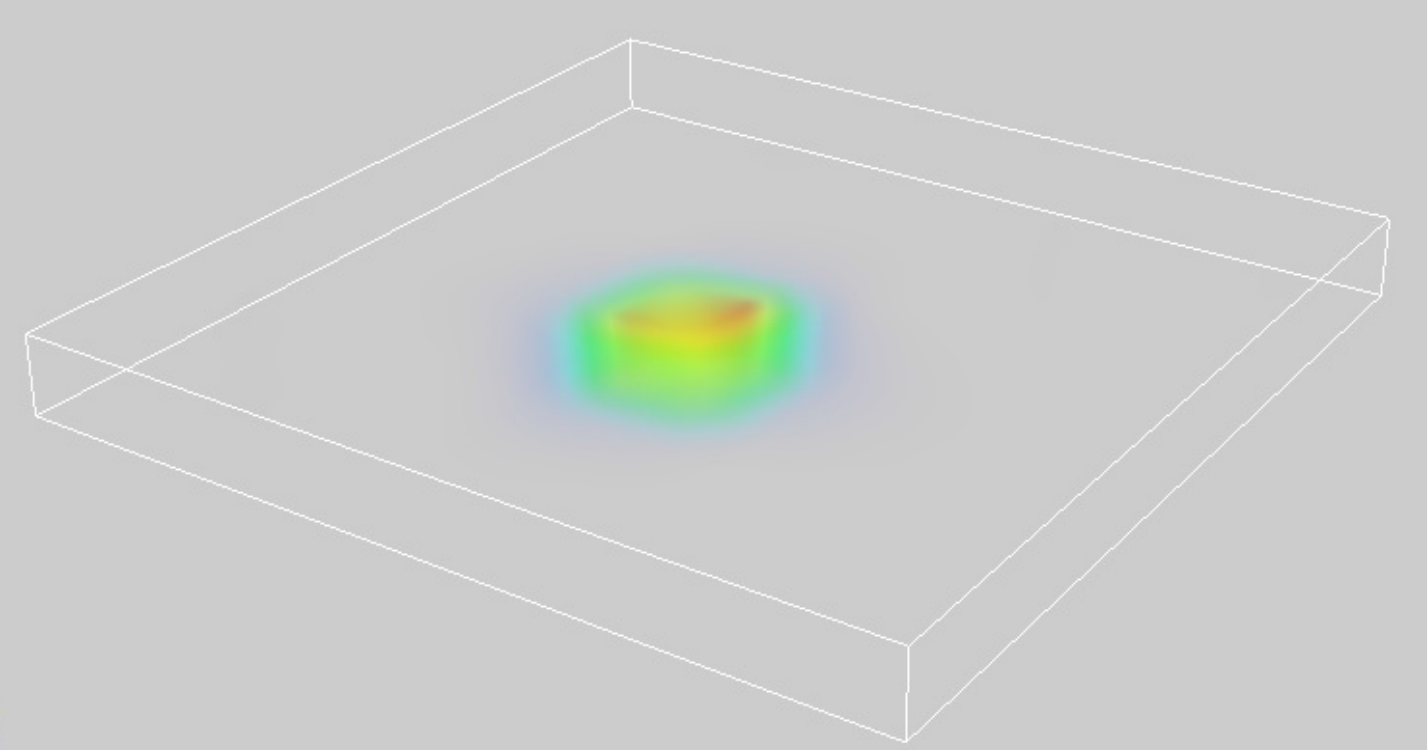}
\vspace{-2mm}
\caption{
Spatial profile of typical wave functions in WTI nano-films
in the presence of moderate disorder ($W=2$).
Panels (a) , (b): Contrasting behaviors in the cases of $N_z$ odd and even.
(a) When $N_z$ is odd ($N_z=3$), the wave function is extended over the entire side surfaces.
(b) When $N_z$ even ($N_z=2$) the wave function is localized around one corner of the prism.
Panels (c), (d): the magnetic flux introduced to examine the spectral flow
shown in Fig. \ref{film} induces also a bound state surrounding
the flux tube piercing the bulk WTI.
(c) $N_z=3$, $\phi = 0.7\pi$; (d) $N_z=2$, $\phi = 0.9\pi$.
}
\label{wf_film}
\end{figure*} 

\section{WTI nanofilm: case of even vs. odd number of atomic layers}

The specificity of the WTI surface states is that
it has ``dark'' surfaces (gapped surfaces).
\cite{dark}
Below
we will make the best use of this existence of the dark surfaces
in the study of the robustness of edge state network against disorder.
The transport character of a WTI thin film shows
also a peculiar 
dependence (an even/odd feature) on the number of stacked atomic layers.
Here, we focus on a rectangular nano-flake
geometry as depicted in Fig. \ref{schema} (a),
and quantify such an even/odd feature
in spectrum
and in the behavior of the wave function in the presence of disorder. 

\subsection{Response to disorder, response to the flux}

A standard way to quantify the response of the system to disorder is
to study the conductance in an open geometry. \cite{KOI}
But here, 
since our primary purpose is to examine the robustness of the edge state {\it network} 
against disorder,
we have chosen to consider a finite, closed system with 
circulating
quasi-1D surface states,
and study their response to disorder by 
examining the spectral flow when an external magnetic flux is introduced piercing the system.
Such an approach can be more straightforwardly adapted to the case of the edge state network.
A non-trivial spectral flow is a smoking gun for the existence of a robust surface state,
and is protected by the Kramers degeneracy, an immediate consequence of the time-reversal symmetry.

Note that 
our bulk material is in a WTI phase with
specific weak indices: $\bm\nu=(0,0,1)$;
i.e.,
surfaces normal to the $z$-direction (the top and bottom surfaces) are dark/gapped surfaces,
so that
the flux inserted (in the $z$-direction) through such surfaces
does not touch the surface states emergent on side surfaces of the system
(see Fig. \ref{schema}).
The role of the flux is then
to twist the boundary condition 
applied to such helical surface states
circulating around the flux
\cite{strong_side, Bjoern}
The flow of the spectrum in the presence of disorder
as a function of the flux
encodes information on the response of the surface states against disorder;
{\rm i.e.}, whether the surface states are localized or extended.

In a WTI nano-flake as depicted in Fig. \ref{schema} (a)
electrons in the mid-gap states are confined onto side surfaces,
forming a closed 1D circuit
consisting of a pair of helical modes.
The electronic properties of the helical states
are governed by the thickness $N_z$ of the flake.
The even/odd feature in the clean limit is discussed 
in the next subsection.
In the presence of disorder, 
such a difference in the behavior of surface 1D modes
is further accentuated.
When $N_z$ is odd,
the 1D modes remain to be perfectly conducting;
while, when $N_z$ is even, 
all the 1D channels tend to get localized.
These two contrasting behaviors can indeed be triggered
by our diagnosis
focusing on the flow of the spectrum
as a function of the flux inserted.

Each eigenstate on the 1D circuit
is characterized by a 
component of momentum $k_\parallel$
along the circumferential direction
[see also part B, where
$k_\parallel$ is explicitly introduced in the formulation given there].
Naturally,
$k_\parallel$ is discretized, 
reflecting the finite circumference of the nano-flake
[see Eqs. (\ref{apbc}) and (\ref{kn})]
so that the energy spectrum is also discretized.
The flux $\Phi$ is here
inserted through a single plaquette
centered roughly at the position of the axis of the prism [see Fig. 1(a)].
An electron in
the 1D surface state circulating the prism 
``feels'' the flux inserted through the Aharonov-Bohm (AB) effect,
provided, of course, that such an (extended) state is existent;
in the presence of disorder
this happens only in the case of $N_z$ odd.
Introduction of the flux results in the shift of 
$k_\parallel$ by an amount of
$\phi = 2\pi (\Phi /\Phi_0)$ with $\Phi_0=h/e$.
As varying $\phi$
over one cycle of AB oscillation: $\phi\in [-\pi,\pi]$,
the discretized energy eigenvalues are interpolated
and reconstructed; see the arguments given 
in the next subsection.
The above arguments can be generalized,
and holds also true in
the presence of disorder.\cite{Buttiker}

Fig. \ref{film} (a-d) are examples of the calculated spectral flow.
Panels (a), (b) deal with the clean limit, while panels (c), (d)
correspond to the disordered case with
$W=2$ (hereafter $W$ is measured in units of $m_{2\parallel}$).
They both demonstrate the contrasting behaviors
in the $\phi$-dependence of the spectrum: $E(\phi)$
in the cases of $N_z$ odd vs. even.
In panels (a), (b) of Fig. \ref{wf_film},
the contrasting behavior of
typical surface wave functions are shown:
when $N_z$ is odd [case of panel (a)],
the surface wave function is extended and
covers the entire side surfaces. 
On contrary,
when $N_z$ is even [case of panel (b)], 
the surface wave function is localized
in the vicinity of one corner of the prism.
In the spectral shown in Fig. \ref{film}
one can also recognize a branch of spectrum
due to bound states formed along the flux introduced
[see also 
Fig. \ref{wf_film} (c), (d)].

\subsection{Nature of the electronic states on WTI surfaces}

To have further insight on the contrasting behaviors in spectral flow
revealed in Fig. 2
in the cases of $N_z$ odd and even,
we start by analyzing this issue based on
the effective theory for WTI surface states.
Such an effective theory has been employed in the analyses of
Refs. \onlinecite{Mong, Fu, Morimoto, Hide},
while more recently it has been explicitly derived from the bulk
effective Hamiltonian.
\cite{arita}
The central ingredients of the theory are two Dirac cones 
that appear in the surface BZ for a side surface
that is parallel to $\bm \nu$.

In the following simulations 
we choose the parameters as in Eq. 
(\ref{param})
to realize a WTI with indices $\bm \nu = (0,0,1)$.
We then put this system into a geometry as shown in Fig. 1 (a),
{\rm i.e.},
a prism of height  $N_z$
and with a constant cross sectional area of size $N_x\times N_y$.
As a result,
surface electronic states
associated with the two Dirac cones
that appear on side surfaces of the prism
are regrouped into those of sub-bands.
The entire spectrum takes the following form;
see Appendix for its justification:
\begin{equation}
E=\pm\sqrt{(A_z \sin{q})^2+(A_\parallel k_\parallel)^2}
\equiv E^\pm(q, k_\parallel),
\label{Eqk}
\end{equation}
where both (i) $q$ and (ii) $k_\parallel$ take discrete values
due to quantization associated, respectively, with
(i) confinement of the surface wave function into a width of $N_z$,
and
(ii) the circular motion around the prism 
of circumference $\xi \simeq 2(N_x+N_y)$.
Naturally, the effect of (i) $N_z$-quantization is more important,
since here we have in mind a situation in which
$N_z \ll \xi$. 

Let us note that
the following arguments 
[and the arguments leading to Eq. (\ref{Eqk})]
are based on an analysis of an idealized circular system,
and is not entirely justified in the rectangular system which is employed in the numerical study. 
Yet, as we see below,
features deduced from the analysis of this idealistic circular model 
are shown to be indeed useful in the interpretation of numerical results 
in the rectangular system, 
to a degree which allows for a quantitative comparison
(Sec. III-C and Sec. III-D).
The underlying hypothesis is that the physics triggered here is quasi-1D
so that the essential features encoded in the spectral flow is well explained 
by the idealized circular model.

Let us now focus on the quantization of $q$.
This $N_z$-quantization regroups the spectrum represented by Eq. (\ref{Eqk}) 
into sub-bands specified by a band index ${m}$ such that
\begin{equation}
E^\pm(q_{{m}}, k_\parallel)\equiv E^\pm_{{m}}(k_\parallel),
\end{equation}
{\rm i.e.},
$q$ has been quantized as 
\begin{eqnarray}
q= {{m}\pi\over N_z+1} \equiv q_{{m}},
\label{qm}
\end{eqnarray}
where $2{m}$ is an integer,
or ${m}=0, \pm {1\over 2}, \pm 1, \pm {3\over 2}, \pm 2, \cdots$.
Yet, an essential observation is here to be added.
Depending on the parity of $N_z$,
not all the values of 
$q$ in Eq. (\ref{qm}) are allowed:
\begin{enumerate}
\item 
If $N_z$ is odd, 
the allowed values of ${m}$ 
in Eq. (\ref{qm}) are restricted to integers:
${m}=0, \pm 1, \pm 2, \cdots, \pm{N_z-1 \over 2}$.
Since $q$ appears squared in Eq. (\ref{Eqk}), or more explicitly,
\begin{eqnarray}
E^\pm_{{m}}(k_\parallel)
=\pm\sqrt{\left(A_z \sin\left({{m}\pi\over N_z+1}\right)\right)^2+(A_\parallel k_\parallel)^2},
\label{en_odd}
\end{eqnarray}
the ${m}=0$ sub-band is non-degenerate, while all the remaining sub-bands are
doubly degenerate:
$E^\pm_{{m}}(k_\parallel)=E^\pm_{-{m}}(k_\parallel)$.
Note that the non-degenerate ${m}=0$ sector represents a linearly dispersing, gapless sub-band:
\begin{equation}
E^\pm_0=\pm A_\parallel k_\parallel,
\end{equation}
while the remaining degenerate sub-bands are all gapped.
\item
If $N_z$ is even, 
${m}$ in Eq. (\ref{qm}) is an {\it half odd} integer; 
${m}=\pm{1 \over 2}, \pm{3 \over 2}, \cdots, \pm{(N_z-1) \over 2}$.
This signifies, in contrast to the $N_z$ odd case,
all the sub-bands are without exception doubly degenerate: 
$E^\pm_{{m}}(k_\parallel)=E^\pm_{-{m}}(k_\parallel)$.
Since these degenerate sub-bands are all gapped,
the entire spectrum is also gapped.
The bottom of the lowest-energy sub-band is located at
\begin{eqnarray}
E^+_{1 \over 2}(k_\parallel =0) = 
 A_z\sin\left({\pi\over 2(N_z+1)}\right).
\label{en_1}
\end{eqnarray}
\end{enumerate}

The second source of the quantization is 
(ii) the circular motion around the prism,
which is applied to discretization of $k_\parallel$.
Here, let us take account of also the effect of the flux
inserted.
Then, the periodic boundary condition associated with
the circular motion is twisted by two types of AB effect: extrinsic and intrinsic.
The extrinsic effect is due to the flux $\phi=2\pi(\Phi/\Phi_0)$,
while the intrinsic effect refers to the Berry phase $\pi$
associated with the so-called spin connection.
\cite{Vishwanath_PRB, Vishwanath_PRL, BBM, Mirlin, disloc, cylindrical}
In any case, the boundary condition associated with
the circular motion is given by
\begin{equation}
e^{i(k_\parallel\xi-\phi)}=-1,
\label{apbc}
\end{equation}
where $\xi$ is 
the circumference of this orbital motion. 
Eq. (\ref{apbc}) determines the quantization rule for $k_\parallel$,
which reads
\begin{equation}
k_\parallel={2\pi\over\xi}\left(n-{1\over 2}+{\phi\over 2\pi}\right)
\equiv k_n(\phi).
\label{kn}
\end{equation}
Generally, introduction of a flux breaks time reversal symmetry of the system.
Only at $\phi=0$ and at $\phi=\pi$
the symmetry remains to hold, 
implying that all the states
at these values of $\phi$ are two-fold degenerate
(Kramers degeneracy).
Since
\begin{equation}
k_n(0)={2\pi\over\xi}\left(n-{1\over 2}\right)= -k_{-n+1}(0)
\end{equation}
and $k_\parallel$ appears squared in Eq. (\ref{Eqk}),
a pair of circular modes with
$k_\parallel =k_n$ and $k_\parallel =k_{-n+1}$
are Kramers partners at $\phi=0$:
\begin{equation}
E^\pm_{{m},n}(\phi=0)=E^\pm_{{m},-n+1}(\phi=0),
\end{equation}
where
\begin{eqnarray}
&&E^\pm_{{m},n}(\phi) \equiv E^\pm(q_{{m}}, k_n(\phi))
\nonumber \\
&=&\pm\sqrt{\left(A_z \sin{
q_{{m}}
}\right)^2+\left(A_\parallel {2\pi\over\xi}\left(n-{1\over 2}+{\phi\over 2\pi}\right)\right)^2}.
\label{Emn}
\end{eqnarray}
The two partners evolve, however,
differently on introduction of $\phi$.
At $\phi=\pi$,
both $n$th and $(-n+1)$th modes find a new partner:
\begin{eqnarray}
k_n(\pi)&=&{2\pi\over\xi}n= -k_{-n}(\pi),
\nonumber \\
k_{-n+1}(\pi)&=&{2\pi\over\xi}(-n+1)= -k_{n-1}(\pi),
\label{k_ex}
\end{eqnarray}
{\rm i.e.},
\begin{eqnarray}
E^\pm_{{m},n}(\pi)=E^\pm_{{m},-n}(\pi),\ \ \
E^\pm_{{m},-n+1}(\pi)=E^\pm_{{m},n-1}(\pi).
\label{en_new}
\end{eqnarray}
Indeed, all the Kramers pairs at $\phi=0$
change their partners as $\phi$ evolves from 0 to $\pi$,
and as argued in Ref. \cite{KN}
this change of the partner is the origin of
a characteristic 
spectral flow $\{E_j (\phi)\}$.
Here,
$\{E_1 (\phi), E_2 (\phi), E_3 (\phi), \cdots \}$
is an energy spectrum at a given value of $\phi$,
with energy eigenvalues
$E_1 (\phi), E_2(\phi), E_3(\phi), \cdots$
sorted in the increasing (or decreasing) order. 
A spectral flow $\{E_j (\phi)\}$
is the entire image of the trajectories of such a set of eigenvalues
when $\phi$ is varied over one cycle of AB oscillation,
$\phi\in [-\pi,\pi]$.
In Fig. 2 and in the subsequent figures
only  half of the flow is shown,
since here $E(-\phi)=E(\phi)$ is guaranteed by time reversal symmetry
of the original model.
In Eqs. (\ref{k_ex}) and (\ref{en_new})
the case of ${m}=n=0$ needs a separate consideration.\cite{dark}
The following relation holds: 
\begin{equation}
E^+_{0,0}(\pi)=E^-_{0,0}(\pi)
\end{equation}
instead of Eq. (\ref{en_new}).

Let us focus on the spectral flow shown in Fig. \ref{film}. First 
recall that 
the spectrum is doubly degenerate at $\phi=0$ and at $\phi=\pi$,
and this degeneracy is ensured by the Kramers theorem.
This holds true both in the clean limit [panels (a) and (b)] and in the presence of disorder [panels (c) and (d)]. 
In the case 
of $N_z$ even,
additional degeneracies occur at an intermediate $\phi$ [see panel (b)],
and these crossings are not protected.
In the presence of disorder
such accidental degeneracies are indeed lifted [panel (d)].
When $N_z$ is odd, 
typically a single ${m}=0$ non-degenerate subband appears
in the relevant low-
energy regime;
such a situation is
indeed predominant in panels (a), (c).
Then, the spectral flow is free from accidental crossings
as mentioned above.
Generally, degenerate subbands with ${m}\neq 0$
may also appear in a relatively high-energy 
region and 
be superposed on top of ${m}=0$ non-degenerate subband.
However, mixing with such pseudo two-fold degenerate subbands 
does not destroy the non-trivial spectral flow.
Here, non-trivialness
refers to the fact that the spectral flow is a connected line 
traversing the entire gap region 
as shown in the case of panel (c).
The reason why this is so is essentially due to
the same logic leading to the $\mathbb{Z}_2$ classification of 2D QSH states. \cite{KaneMele_Z2}
In the case 
of $N_z$ even,
anti-crossings at an intermediate $\phi$ make the spectral flow {\it trivial}, 
{\rm i.e.}, the spectrum consists of disconnected lines.

\begin{figure*}[htbp]
(a)
\includegraphics[width=70mm, bb=0 0 260 262]{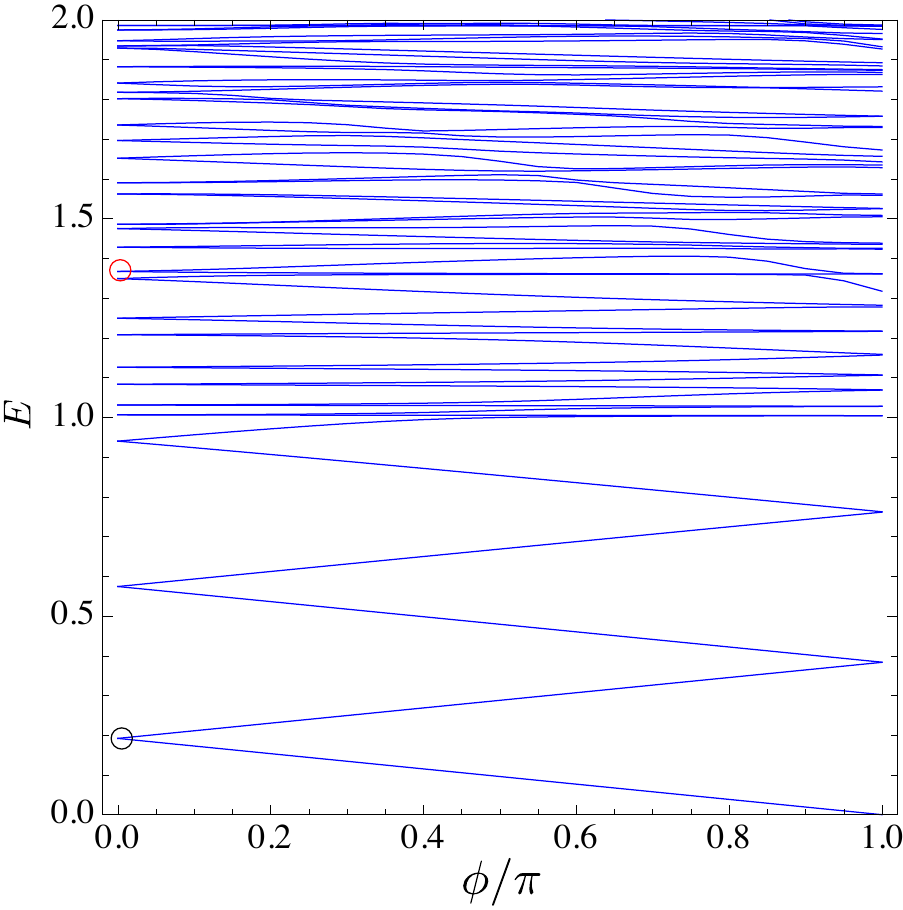}
(b)
\includegraphics[width=70mm, bb=0 0 260 260]{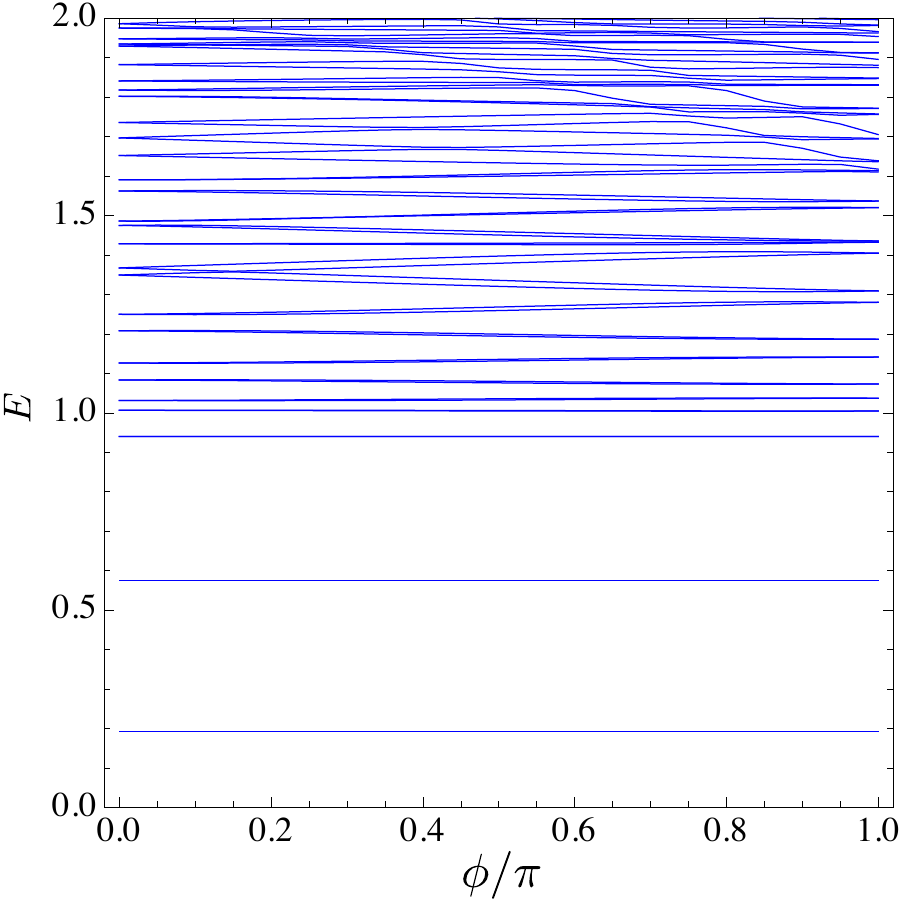}
\\
(c)
\includegraphics[width=80mm, bb=0 0 430 173]{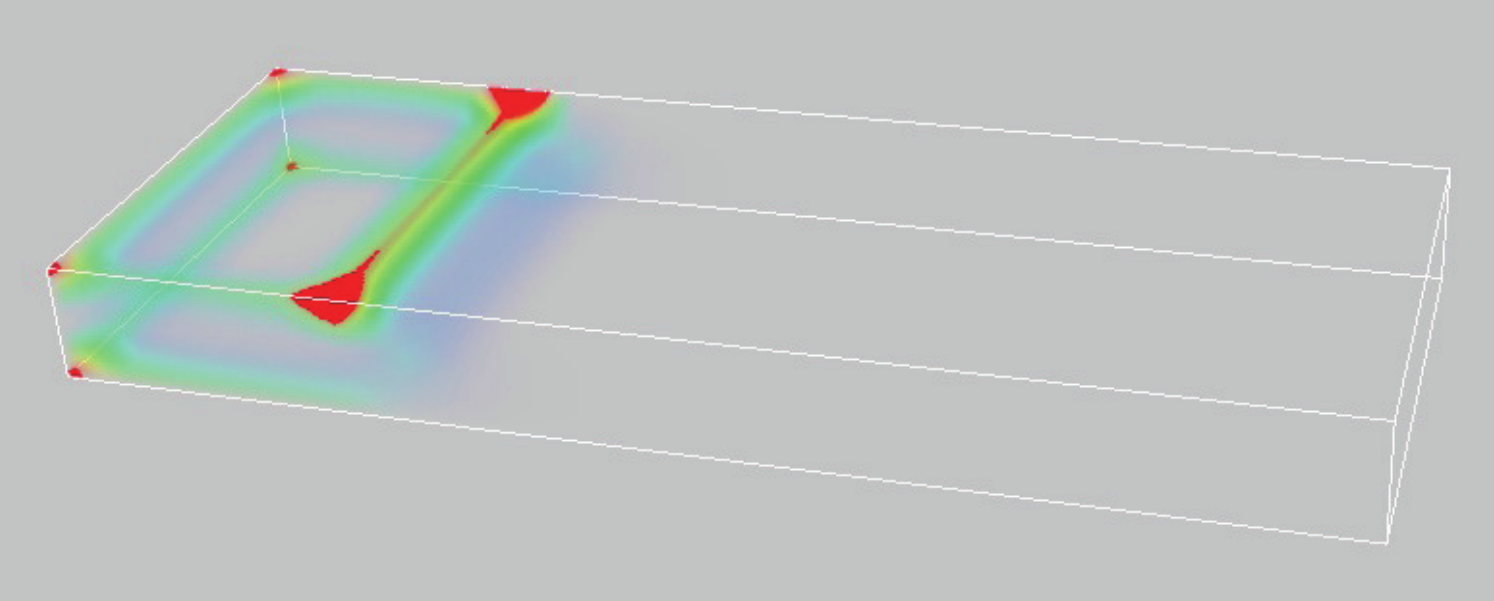}
(d)
\includegraphics[width=80mm, bb=0 0 423 170]{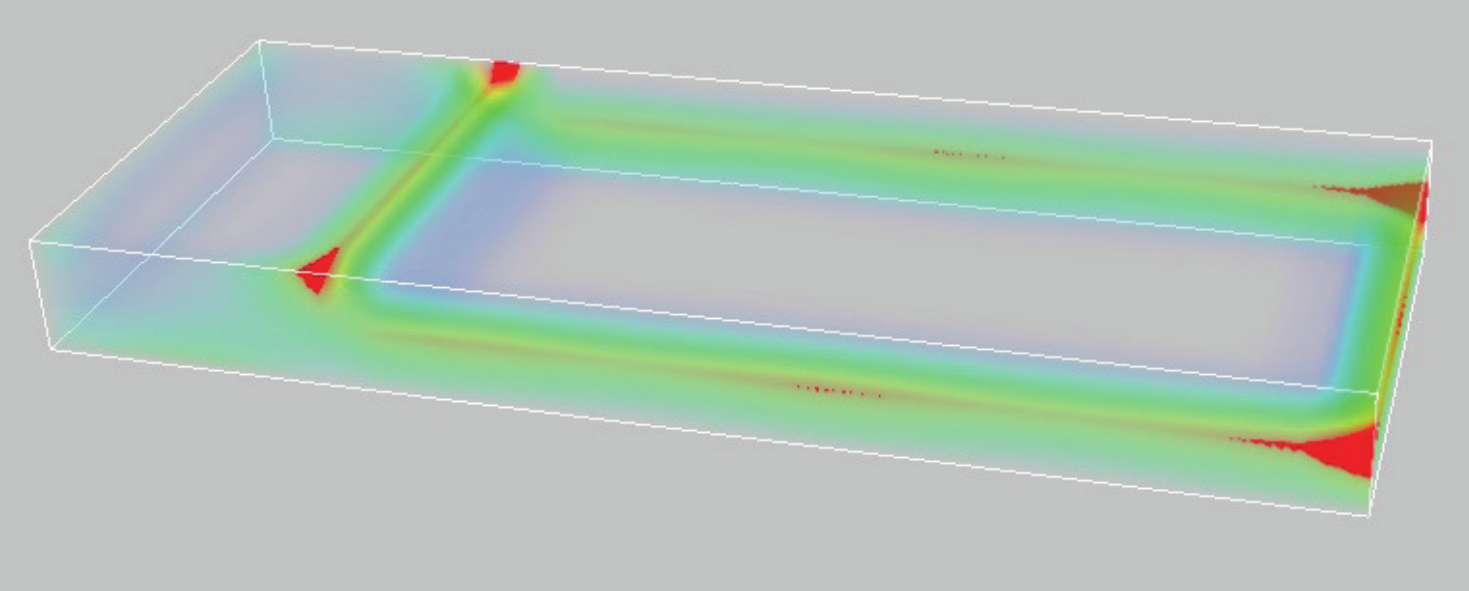}
\vspace{-2mm}
\caption{Case of a simple step:
$N_1=3$, $N_2=2$.
Evolution of the spectrum when a flux is introduced: 
(a) $E(\phi_1)$ with $\phi_2=0$, 
(b) $E(\phi_2)$ with $\phi_1=0$. 
The flux is inserted as in Fig. \ref{schema} (b) 
in the clean limit of $W=0$. 
The flux is measured in units of
$\phi_1 = 2\pi \Phi_1/\Phi_0$,
$\phi_2 = 2\pi \Phi_2/\Phi_0$.
Spatial profile of the wave function: 
(c) case of the 1st level indicated by a black circle in panel (a), 
(d) case of the 40th level indicated by a red circle in panel (a).
The levels are sorted in the increasing order in $E$
.
}
\label{step1_W0}
\end{figure*}

\subsection{Even/odd features in the spectral flow}

Based on the observations so far established in the light of the surface effective theory,
let us re-examine the spectral flow shown in the four panels of Fig. 2
in more detail.
Panels (a), (c) show a calculated spectral flow 
in the case of $N_z$ odd ($N_z=3$),
while panels (b), (d)
correspond to the case of $N_z$ even ($N_z=2$).
Panels (a), (b) represent a spectral flow in the clean limit, 
while (c), (d) are those of the disordered case: $W=2$.
Model parameters are set as
$N_x=N_y=12$ and $A_\parallel=A_z=2$ (measured in units of $m_{2\parallel}$).

\subsubsection{Case of $N_z$ odd: extended, perfectly conducting}

Let us first focus on 
Fig. \ref{film} (a):
case of $N_z=3$ in the clean limit.
In the range of energies shown in the figure,
the low-lying part of three sub-bands
with ${m}=0$ and ${m}=\pm 1$
of Eq. (\ref{en_odd}),
are relevant,
contributing to the spectral flow.
The ${m}=0$ sub-band is non-degenerate and gapless,
which is responsible for the non-trivialness of the flow.
In the $E>0$ sector,
the bottom of the degenerate
${m}=\pm 1$
sub-bands are located at
\begin{equation}
E^+_1 (k_\parallel =0)= A_z\sin\left({\pi \over N_z+1}\right) 
=2\sin{\pi \over 4}=\sqrt{2}.
\label{en_2}
\end{equation}
The simple zigzag pattern below this threshold energy
is purely due to the ${m}=0$ sub-band,
while above this energy
the two contributions are superposed.

In the pure $m=0$ regime
the pitch of the zigzag pattern is given as
\begin{eqnarray}
\Delta E_{\rm pitch} = A_\parallel{2\pi\over\xi}={\pi\over 12}
\simeq 0.2618
\end{eqnarray}
[see Eq. (\ref{Emn})].
In the presence of disorder 
[Fig. \ref{film} (c)]
this pitch is modified by the mixing of 
${m}=0$ and ${m}=\pm 1$ sub-bands,
while the {\it connectedness}
of the zigzag pattern is maintained;
the spectral flow remains {\it non-trivial}.
Crossing of the spectra at $\phi=0$ and $\phi=\pi$ 
is a consequence of the time reversal symmetry
(Kramers degeneracy),
which is unaffected by introduction of non-magnetic impurities
considered here.
Robustness of the continuous zigzag pattern 
is a clear signature
that a pair of surface helical channels are robust against disorder,
and the corresponding wave function is {\it extended}
despite the presence of disorder
[Fig. \ref{wf_film} (a)].

\subsubsection{Case of $N_z$ even: all the states get localized}

If $N_z$ is even,
the situation is much different.
First, the spectrum is gapped by a finite-size quantization
[see Eq. (\ref{en_1})].
The half width of this gap is 
$E_{1\over 2}^+(k_{\parallel}=0)= 1$
in the present choice of parameters
[{\rm cf.} Fig. 2, panel (b)].
Above this threshold energy,
two pseudo degenerate sub-bands with ${m}=\pm {1 \over 2}$
become available for edge/surface conduction.
In the clean limit [panel (b)]
these two sub-bands form a zigzag pattern somewhat resembling
the case of $N_z$ odd.
Note that the two sub-bands are not completely 
degenerate;
they interfere due to the existence of corners, and as a result
their spectrum repel each other.
Also importantly, there is a crossing once per each period $\phi\in [0,\pi]$
between these pseudo degenerate sub-bands 
at a (non-protected) 
intermediate
value of $\phi$
(recall the arguments in the previous subsection).
Crossings occur with a counter-propagating 
partner,
and between neighboring $k_n$ modes
[see Eq. (\ref{kn})].

Now, as we switch on disorder [see panel (d)],
a crucial difference arises
from the case of $N_z$ odd [panels (c)].
The zigzag is broken apart into many pieces;
the spectral flow is indeed {\it trivial} in this case.
In Fig. 3 (b)
the spatial profile of
the corresponding wave function is shown.
In consistency with the {\it trivial} spectral flow
the wave function is {\it localized} in the vicinity of one corner of the prism.

\begin{figure*}[htbp]
(a)
\includegraphics[width=70mm, bb =0 0 260 260]{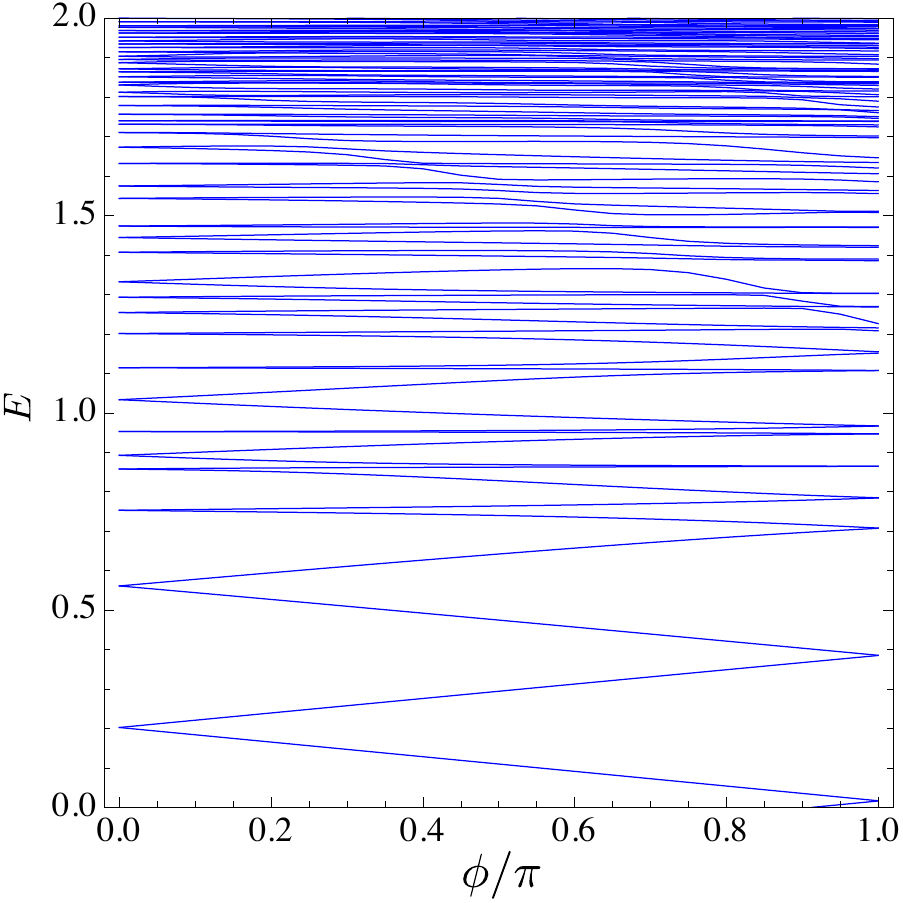}
(b)
\includegraphics[width=70mm, bb =0 0 260 260]{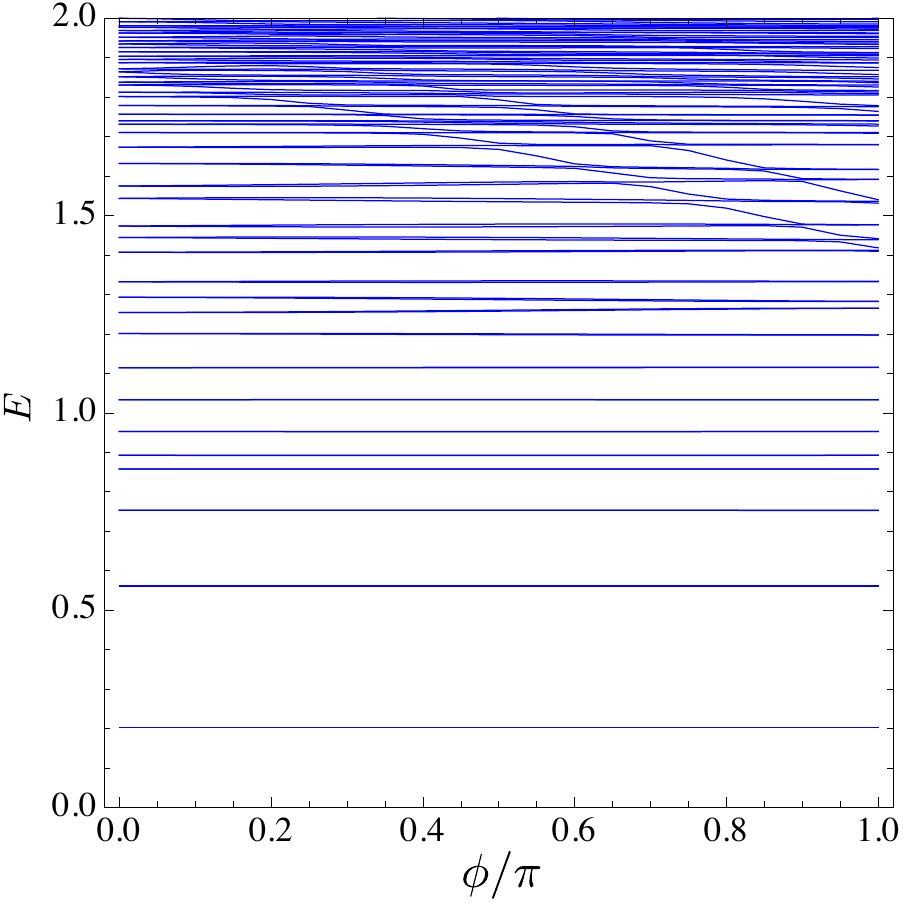}
\vspace{-2mm}
\caption{Similar to Fig. \ref{step1_W0} (a) and (b), 
but in the presence of disorder: $W=2$.
(a) plots of $E(\phi_1)$,
(b) $E(\phi_2)$.
}
\label{step1_W2}
\end{figure*}

\subsection{Bound states}

In the spectral flow shown in the four panels of Fig. \ref{film}, 
one can recognize a separate branch that are superposed 
on top of the spectral flow we have so far focused on.
This separate branch
stems from a 
bound 
state 
induced by the flux insertion.
\cite{disloc, cylindrical, Ran_nphys}
This can be verified explicitly by
inspecting the spatial profile of the corresponding wave function
as shown in Fig. \ref{wf_film} (c), (d). 
The figure indicates that such bound states are localized around a plaquette 
through which the flux is inserted. 
That is, they can be regarded as localized states on the surface of 
a prism-shaped hole ({\rm i.e.} flux tube) corresponding to the plaquette,
where the circumference $\xi_b$ of the hole is $\xi_b \simeq 4a$ 
with $a$ being the lattice constant.
The reason why such bound states
appear in the spectrum can be read from Eq. (\ref{Emn}), 
while here the typical length scale is $\xi_b$
associated with the quantization of $k_\parallel=k_n$
[see Eq. (\ref{kn})]. 
As $a$ is chosen to be unity in the simulation, $\xi_b\simeq 4$.
The fact that $\xi_b$ is on the order of $N_z$ implies that
$q$-quantization and $k_\parallel$-quantization are equally important.
This is contrasting to the case of surface states on 
side surfaces,
in which $N_z \ll \xi$ holds,
indicating that the $q$-quantization is much more important.
In the low-energy regime shown in Fig. \ref{film}
only the $n=0$ (or $n=1$ on the $\phi <0$ side) 
sector is relevant.

When $N_z$ is odd,
$q$-quantization allows for a zero mode: ${m}=0$ in Eq. (\ref{qm}). 
The separate branch that appears in the spectral flow
shown in Fig. \ref{film} (a), (c),
and the spatial profile of the wave function shown in 
Fig. \ref{wf_film} (c)
are due to such a bound state with
${m}=0$ and $n=0$.
Since
\begin{eqnarray}
q_0&=&0,
\nonumber \\
k_0(\phi)&=&{2\pi\over\xi_b}\left(-{1\over 2}+{\phi\over 2\pi}\right),
\end{eqnarray}
the energy of this bound state is zero at $\phi=\pi$.
Taking
$\xi_b\simeq 4$ 
into account,
one can also estimate the rough energy ``dispersion'' $E(\phi)$
of this bound state from Eq. (\ref{Emn})
as
\begin{equation}
E^{(0)\pm}_{\rm bound}
=E^\pm_{0,0} (\phi)
= \pm A_\parallel {\pi-\phi\over 4}.
\label{bound_0}
\end{equation}
In the spectral flow
shown in Fig. \ref{film} (a), (c),
the separate branch due to bound state
shows indeed such a linear dispersion
in the vicinity of the
gap closing
at $\phi=\pi$.
In the high energy part of the spectral flow
in the clean limit [Fig. \ref{film} (a)],
one can also recognize the second sets of bound states,
which are due to ${m}=\pm 1$ and $n=0$.

When $N_z$ is even,
$q$-quantization has no zero mode.
The separate branch that can be seen in Fig. \ref{film} (b), (d)
are due to bound states with
${m}=\pm \frac{1}{2}$ and $n=0$.
The spatial profile of the wave function in this case
is shown in Fig. \ref{wf_film} (d).
From Eq. (\ref{Emn})
one can make a rough estimate of $E(\phi)$
for such 
gapped bound states:
\begin{eqnarray}
E^{({m})\pm}_{\rm bound} (\phi)
=\pm\sqrt{\left(A_z \sin\left({{m}\pi\over N_z+1}\right)\right)^2
+A_\parallel^2 \left({\pi-\phi\over 4}\right)^2},
\label{bound_m}
\end{eqnarray}
where
${m}=\pm {1 \over 2}$ in the present case with $N_z=2$, 
{\rm i.e.}, 
\begin{equation}
E^{({1 \over 2})\pm}_{\rm bound} (\pi) = \pm A_z \sin\left({\pi\over 2(N_z+1)} \right)
= \pm 1.
\label{bound_1}
\end{equation}
Setting ${m}=\pm 1$, one can apply 
Eq. (\ref{bound_m}) to the second excited bound states
in the case of 
$N_z=3$,
\begin{equation}
E^{(1)\pm}_{\rm bound} (\pi) = \pm A_z \sin\left({\pi\over N_z+1}\right)
 = \pm \sqrt{2}.
\label{bound_2}
\end{equation}
Though such estimates as
given in Eqs. (\ref{bound_0}), (\ref{bound_m}), (\ref{bound_1}), (\ref{bound_2})
are very rough ones,
they still show a qualitatively good agreement with
the calculated spectral flow presented
as four panels in Fig. \ref{film}.

\begin{figure*}[htbp]
(a)
\includegraphics[width=80mm, bb=0 0 260 255]{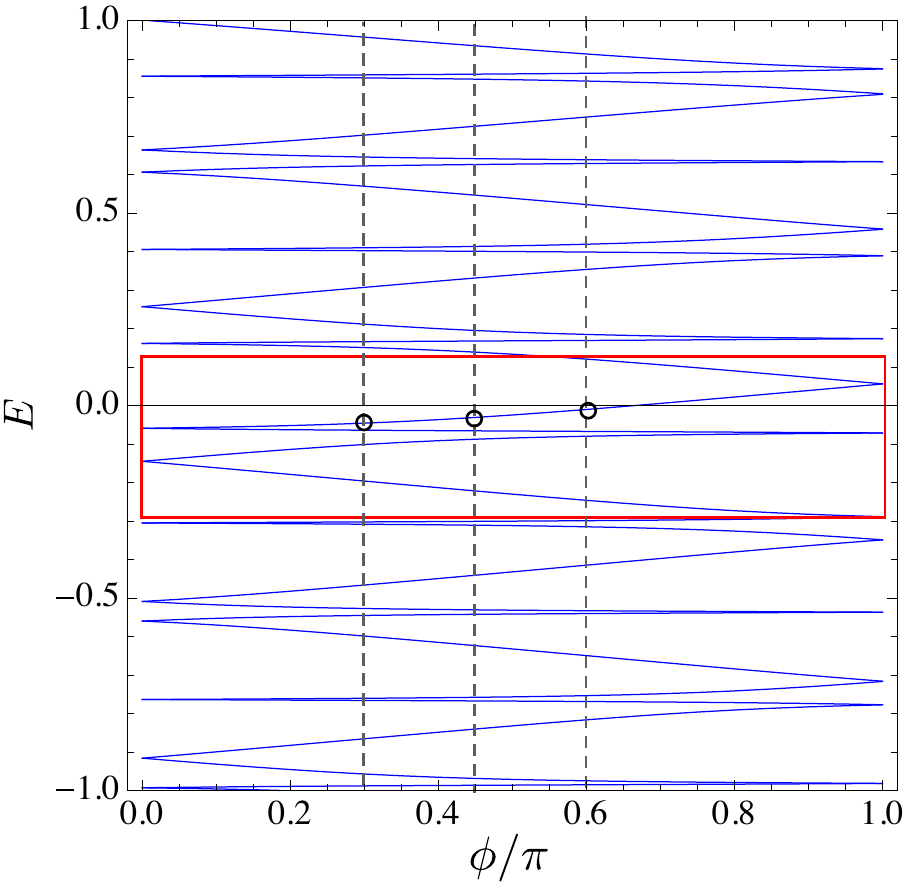}
(b)
\includegraphics[width=80mm, bb=0 0 260 254]{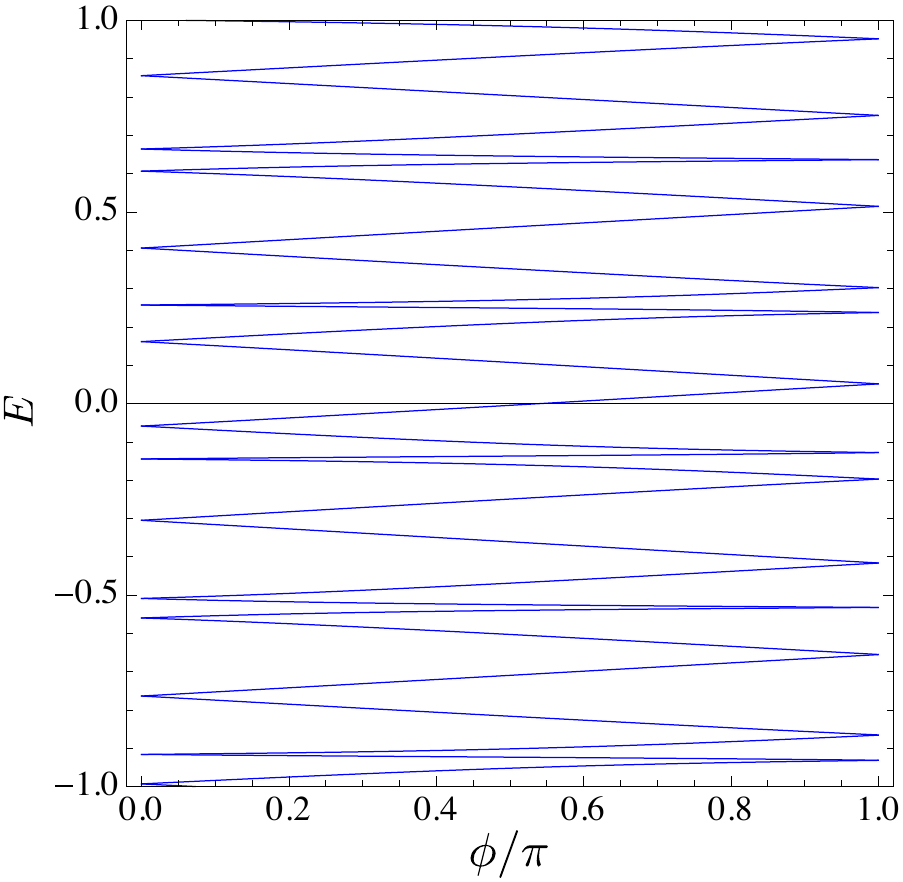}
\vspace{-2mm}
\caption{
Two types of spectral flow $E(\phi)$ in the two-step geometry.
The flux is inserted in two different ways:
$(\phi_1,\phi_2)=(\phi, 0)$ in panel (a);
the flux is inserted on the $N_1$- ($N_z=3$) side,
while
$(\phi_1,\phi_2)=(0, \phi)$ in panel (b);
the flux is inserted on the $N_2$- ($N_z=1$) side.
A moderate strength of disorder is also introduced
($W=2$).
}
\label{step2_W2}
\end{figure*}

\section{Perfectly conducting channels emergent in WTI nano-architectures}

Let us consider the step geometries
as depicted in Fig. 1 (b), (c).
A PCC 
appears
along a step (or steps) etched on the surface of a WTI.
In the case of a single step, the width of an otherwise uniform nano-flake differs 
on the two sides of the step
simply by the height of the step, 
which we assume to be odd. 
Then, the height of the two sides is either odd-even, or even-odd. 
A PCC appears naturally on the side the width of which is odd, 
and 
is smoothly connected to the PCC along the step, 
forming a closed nano-circuit of PCC.
We probe the nature of such PCC
along the step and around the side surfaces
by studying response of the system to flux $\Phi_1$ and $\Phi_2$ 
introduced on either side of the step,
independently.

In the case of the double step
[see panel (c) of Fig. 1], 
provided that the height of two steps 
is both odd, there are two possible options for the width of nano-flake 
on the two sides of the step. 
The height of the two sides is either odd-odd or even-even. 
In the case of the even-even combination
there appears no PCC 
circulating around the nano-flake, so that two PCCs in the step region 
connect with each other, forming a single closed loop.
This means that this combination 
reduces to the previous case of a single nano-circuit.
The odd-odd combination realizes a typical example of multiple nano-circuit 
we intend to highlight in part B of this section.

\subsection{Case of the single step:
a robust perfectly conducting channel running along the step}

The model geometries depicted in Fig. 1 (b), (c) can be regarded
as a set of two rectangular prisms of different height, say, $N_1$ and $N_2$
joined together via side surfaces.
The gapped surfaces are on the top and bottom surfaces.
Here,
we align them such that the bottom surfaces
are smoothly connected (case of the single step).
Then, if $N_1\neq N_2$,
a step of height $\Delta N = N_1 -N_2$ appears on the top surface.
If $\Delta N$ is odd,
there appears a pair of 1D protected helical modes (i.e., PCCs)
along the step,\cite{dark}
while
if $\Delta N$ is even, this is no longer the case;
pseudo 1D modes are gapped out by the finite size effect
and do not appear in the low energy spectrum.
Indeed,
this odd/even feature with respect to $\Delta N$
stems from the difference of
spectrum in the two cases 
[see Eq.(\ref{en_odd})].

Let us consider a situation
in which $N_1$ is odd and $N_2$ is even, say,
$N_1=3$ and $N_2=2$.
Since $\Delta N$ is odd ($=1$),
there appears a pair of 1D protected modes along the step.
\cite{dark}
Here, one can apply the same arguments leading to Eq. (\ref{en_odd})
for surface states emergent at the step.
Naturally, these 1D helical modes
cannot be confined in a finite segment of the step.
They must be extended over to side surfaces of the prism.
Surface states on such side surfaces become 
gapless
[{\rm i.e.},
excepting the $k_\parallel$-discretization due to Eq. (\ref{kn})]
when $N_z$ is odd
[case of Eqs. (\ref{en_odd}) with ${m}=0$].
In the situation we consider, this happens on the $N_1$-side
[Fig. \ref{schema} (b)].
Thus, in the region of $E>0$ but 
below 
the bottom of the surface sub-band on the $N_2$-side
located at the energy given by Eq. (\ref{en_1}), 
the 1D modes along the step form a closed loop solely with
the pseudo 2D surface modes on the $N_1$-side 
[see Fig. \ref{step1_W0} (c)].
Above this threshold energy 
({\rm i.e.}, the bottom of the sub-band on the $N_2$-side)
an electron propagating 
via a 1D channel along the step and incident at a quantum junction that 
appears
at the end of the step can turn
either to the $N_1$- or to the $N_2$-side 
[see Fig. \ref{step1_W0} (d)].
At energies $E$ above
the bottom of second sub-band on the $N_1$-side, 
given by 
Eq. (\ref{en_2}),
there seem to be {\it a priori}
three, two and one pair of channels, incident, respectively,
from the $N_1$-, $N_2$- and the step sides 
to the quantum junction.

In the presence of disorder, however,
not all of these channels survive.
To see the robustness of different channels against disorder,
here, we have studied the spectral flow in the system
under insertion of a magnetic flux 
in two different configurations.
In Figs. \ref{step1_W0} and \ref{step1_W2}
such a spectral flow is presented
in panel (a)
under a flux configuration of $(\phi_1,\phi_2)=(\phi, 0)$,
where
$\phi_1=2\pi(\Phi_1/\Phi_0)$ and
$\phi_2=2\pi(\Phi_2/\Phi_0)$
represent, respectively,
a flux inserted on the 
$N_1$-side 
and
on the $N_2$-side. 
In panel (b) of Figs. \ref{step1_W0} and \ref{step1_W2}
a different configuration of $(\phi_1,\phi_2)=(0, \phi)$
is studied.
Two panels of
Fig. \ref{step1_W0} represent a spectral flow
in the clean limit ($W=0$),
while
a moderate strength of disorder ($W=2$)
is introduced in the examples shown in Fig. \ref{step1_W2}.
As one can clearly see 
in Fig. \ref{step1_W2}
a nontrivial spectral flow is still persistent in panel (a), 
{\rm i.e.}, the spectral flow is {\it non-trivial} in this case,
while
the flux dependence is almost extinct in panel (b);
the spectral flow is {\it trivial} in this case. 
This indicates 
that in the single step geometry as depicted in Fig. 1 (b) 
with $N_1$ odd ($=3$), $N_2$ even ($=2$) 
[and therefore, $\Delta N$ odd ($=1$)],
an electron in the 1D protected channel at the step is {\it selectively} transmitted to 
the $N_1$-side in the presence of disorder: $W\neq 0$.

\begin{figure*}[htbp]
(a)
\includegraphics[width=70mm, bb=0 0 260 247]{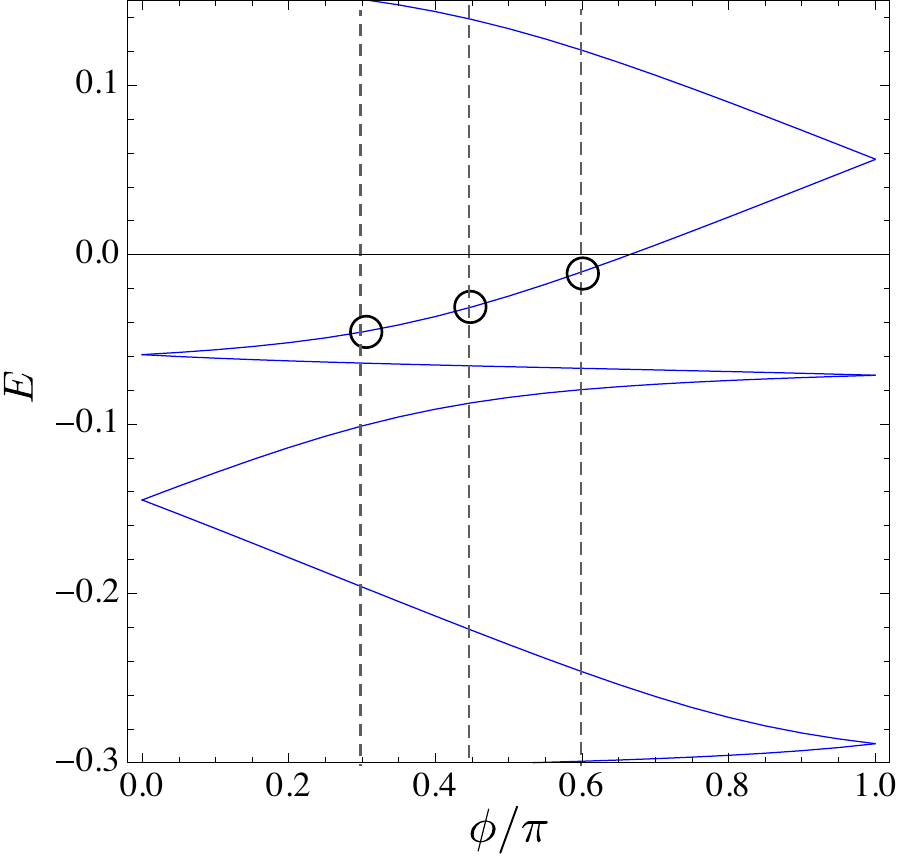}
(b)
\includegraphics[width=70mm, bb=0 0 310 281]{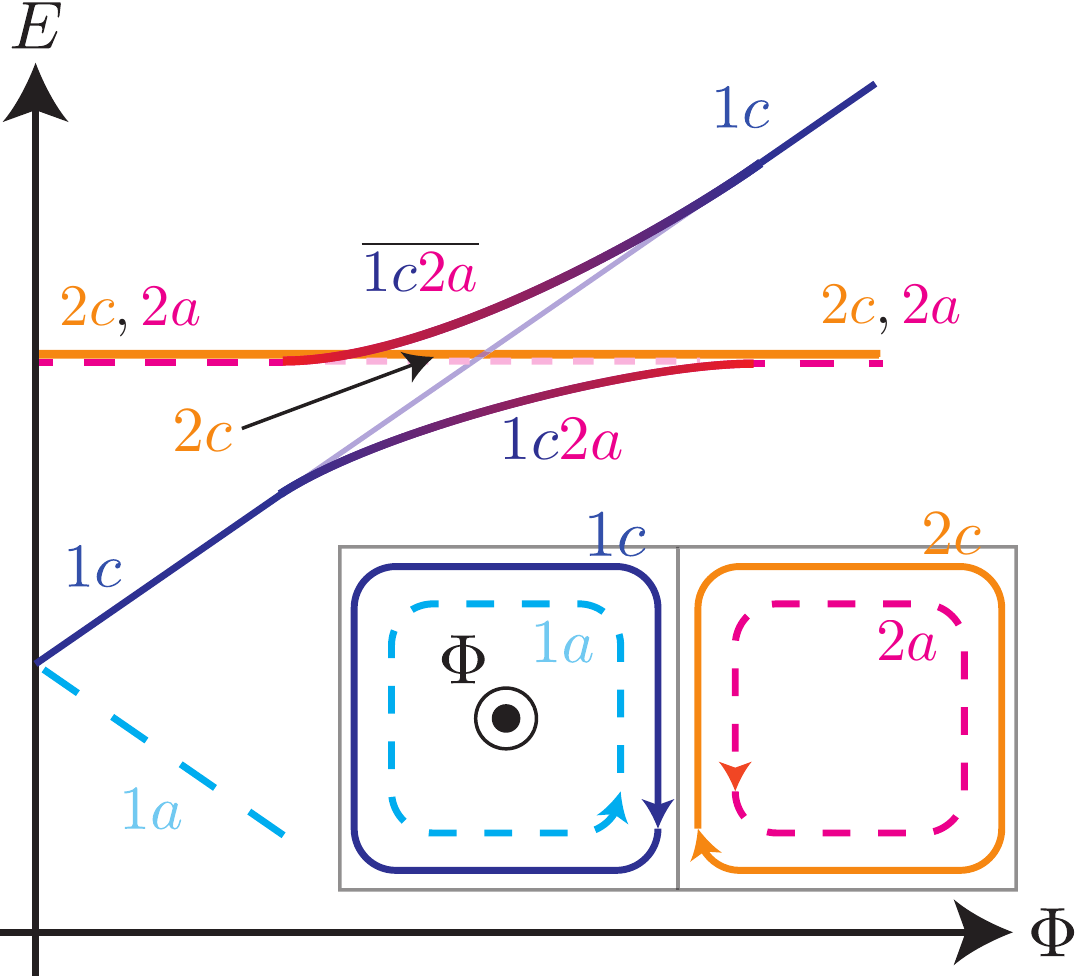}
\vspace{-2mm}
\caption{(a) Zoom-up of the spectrum;
Fig. \ref{step2_W2}, panel (a) the part in red frame.
(b)
Schematic interpretation of panel (a).}
\label{step2_zoom}
\end{figure*}

\subsection{Two step geometry:
an even number of channels running in parallel}

In the previous examples,
only surfaces consisting of an odd number of channels 
are robust against disorder, 
and such surfaces occur when the layer number $N_z$ is odd. 
A natural question that arises here is what happens
in a geometry as depicted in Fig. \ref{schema} (c)
if $N_1$, $N_2$, $\Delta N_1$, $\Delta N_2$ are all odd?
We consider typically the case of 
$N_1=3$ and $N_2=1$ with $\Delta N_1 =1$ and $\Delta N_2 =1$.

Fig. \ref{step2_W2} shows response of such a system against
flux insertion both on the $N_1$- and $N_2$-sides
[see configuration of the two types of flux insertion
$\Phi_1$ and $\Phi_2$ in Fig. \ref{schema} (b)
].
The two panels of
Fig. \ref{step2_W2} show evolution of the spectrum at $W=2$
as a function of the flux $\phi$
when the flux is inserted either 
on the $N_1$- or on the $N_2$-side;
$(\phi_1, \phi_2) = (\phi, 0)$ in panel (a), and
$(\phi_1, \phi_2) = (0, \phi)$ in panel (b).

We begin by pointing out 
two specific features that can be seen 
in the two panels of Fig. \ref{step2_W2}.
First, both panels show a non-trivial spectral flow, in which
each member of a Kramers pair at $\phi=0$ 
changes its partner at $\phi=\pi$.
This means that there exists a robust perfectly conducting channel
both around the prism 1 and around the prism 2.
Provided that
$\Delta N_1$ and $\Delta N_2$ are both odd
($\Delta N_1=\Delta N_2 =1$),
one expects {\it a priori}
an even number ($=2$) of channels 
running in parallel at the connection of two prisms.
Yet, the obtained spectral flow implies that
the two channels behave
{\it as if there were an odd number of channels}.
We will clarify this point later.
Secondly,
the two spectra are complementary in the sense
that those partners that are sensitive to $\Phi_1$
are insensitive to
$\Phi_2$, and vice versa.
Note that
at $\phi=0$ the two plots coincide, since the two simulation is done 
for the same configuration of impurities.

Those states that are sensitive to the flux in 
Fig. \ref{step2_W2} (a) stem from states that goes around the prism 1
[1c and 1a modes in Fig. \ref{step2_zoom}, panel (b)],
while
those which are insensitive 
to the flux are states that goes around the prism 2
(2c and 2a, {\it ibid.}).
To check these assumptions,
here, we have designed the system such that
the circumference $L_2$ of 2c and 2a modes
is twice as long as that of 1c and 1a ($=L_1$): $L_2 \simeq 2 L_1$.
Note that
separation of the levels associated with a 1D channel due to finite size
is inversely proportional to its circumference.
In panel (a) those pairs that are sensitive to the flux
are spaced by a distance twice as large as those
which are insensitive to the flux.
Since the former is assumed to stem from 1c and 1a,
while the latter from 2c and 2a,
this makes perfectly sense.

\begin{figure}[htbp]
(a)
\includegraphics[width=80mm, bb=0 0 523 244]{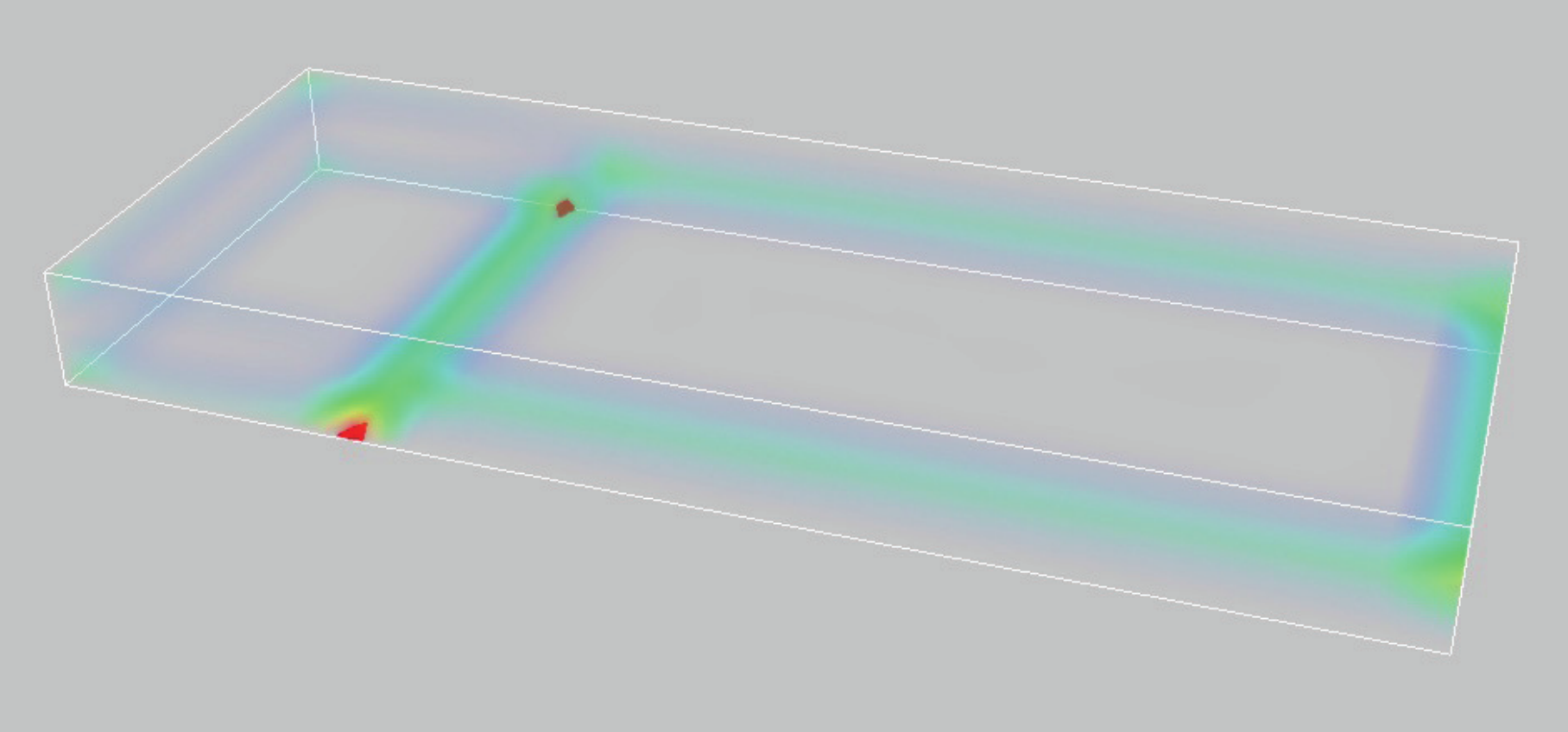}
(b)
\includegraphics[width=80mm, bb=0 0 515 245]{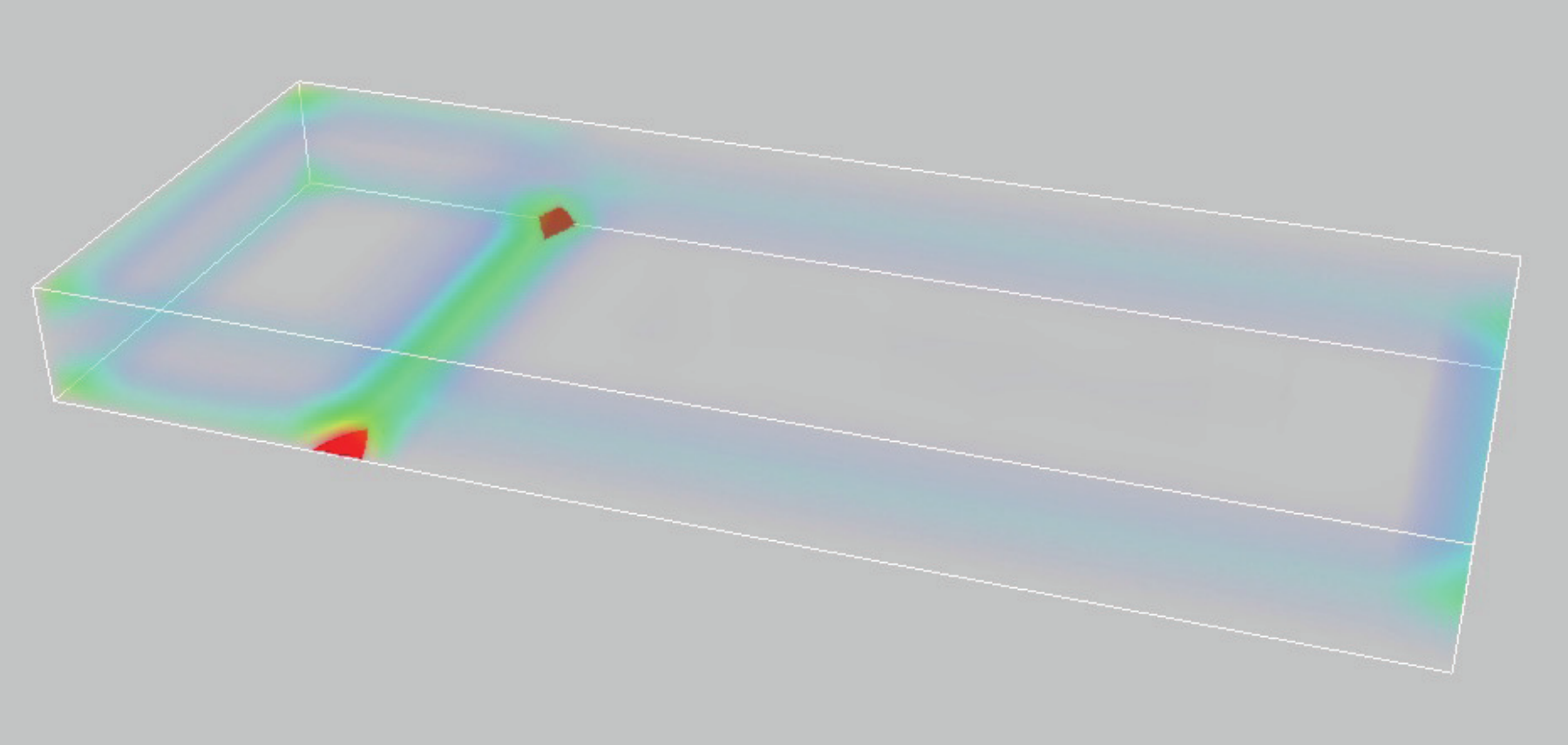}
(c)
\includegraphics[width=80mm, bb=0 0 518 248]{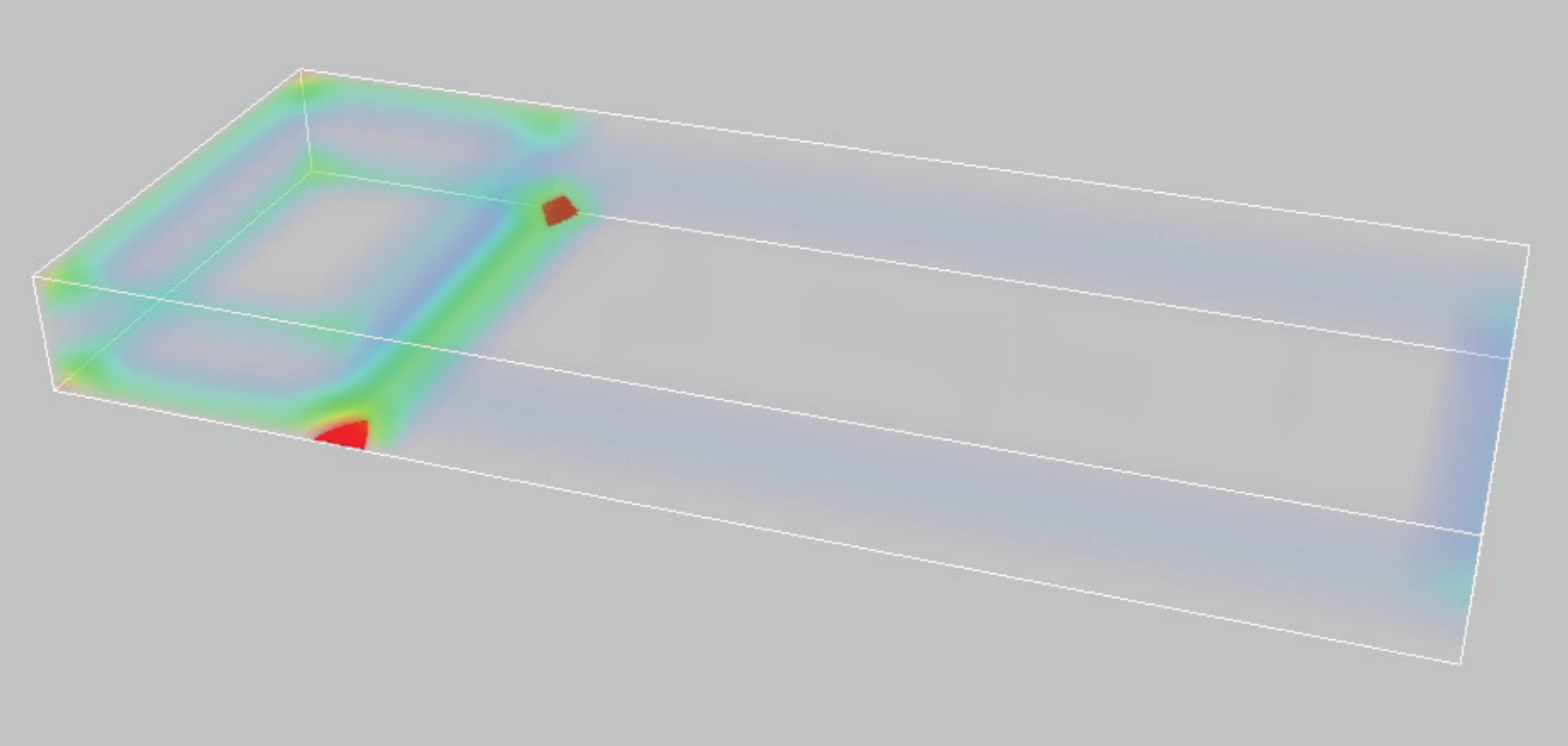}
\vspace{-2mm}
\caption{Evolution of the $|\overline{1c2b}\rangle$ 
state: snap shots of the wave function
at
(a) $\phi/\pi=0.3$, (b) $\phi/\pi=0.45$, (c) $\phi/\pi=0.6$.}
\label{wf_step2}
\end{figure}

These being said, let us come back to the question:
why are the even number of channels incident at the steps
robust against disorder?
Why do they behave like an odd number of channels,
showing a nontrivial spectral flow?
Our short answer is the following:
if we focus on some energy $E$,
say, in the spectrum shown in Fig. \ref{step2_W2} (a),
there exists indeed an odd number of (here, only one)
channel(s) at the step.
To elaborate
what this actually implies,
let us divide the spectrum in energy
into two types of pieces.
The first type of pieces are sensitive to the flux $\phi$,
while the remaining pieces
are insensitive.
The first type of pieces are
due to states which
in real space distributed predominantly in prism 1,
while
the remaining part of the spectrum is due to states
stemming from prism 2.
As varying $\phi$, one can 
evolve
a state incident mainly in prism 1,
into
a state extended over to the side of prism 2
[see Fig. \ref{wf_step2}], 
but then the energy is changed.
Or, conversely, if an energy is given,
there exists a certain value of $\phi$
at which an eigenstate of the system is available.
Then at this energy,
the spatial profile of the available electronic state
is uniquely determined.
When energy is varied,
both electronic states in prism 1 and the ones in prism 2 
become available, but never at the same energy.
As a result,
the spectral flow ``segregates'' into two regions.
The two regions do not coexist, but are connected smoothly.
At a given energy $E$,
there exists only a single state
(at some value of $\phi$)
{\it either} on the $N_1$- or on the $N_2$-side
(or sometimes in between).
This serves as the ``traffic rule'' 
applied here to the T-junctions at both ends of the step.
The signal at the T-junction permits either a left or a right turn depending on 
the energy of the incident electron.

Let us try to formulate
how this {\it segregation} occurs
by zooming up a part of the spectral flow
shown in the red frame of Fig. \ref{step2_W2} (a).
Fig. \ref{step2_zoom} (a)
shows an enlarged image of this part of the flow.
Those parts of the spectrum 
that are sensitive to the flux $\phi_1$
are due to states 
circulating
the prism 1 
either in the clockwise 
or in the anti-clockwise direction
[1c and 1a modes in Fig. \ref{step2_zoom} (b)].
On contrary,
those parts of the spectrum that are insensitive to the flux $\phi_1$
are due to states 
circulating
the prism 2 (2c and 2a modes).
However, in the presence of the step region at which two prisms are coupled, 
the eigenstates of the system become a combination of
1c/1a and 2c/2a states;
they are mixed and recombined in the step region.
In Fig. \ref{step2_zoom} (a)
two Kramers pairs, 1c/1a and 2c/2a, are incident at $\phi=0$ 
apart in energy.
Here, we consider 
the spectral flow of the system under a flux inserted in the configuration
of $(\phi_1,\phi_2)= (\phi, 0)$.
As $\phi_1$ ($=\phi$) is increased, 1c/1a pair breaks; 
$1c$ is the upward branch, which intersects with the 2c/2a pair 
located initially above.
As schematically illustrated in panel (b),
the spectral feature 
in the vicinity of this intersection
can be understood
as the result of anti-crossing between $1c$ and, 
intrinsically,
a linear combination 
of $2c$ and $2a$.
Since the spin quantization axis of  the two helical pairs:
1c/1a and 2c/2a,
are not 
not necessarily 
aligned,
one expects {\it a priori}
recombination of such modes at the step region.
In principle,
$1c$ is coupled to either $2c$ or $2a$, or to both.
Let us introduce the linear combination:
\begin{eqnarray}
|2b\rangle &=& c_{2c} |2c\rangle + c^{\ast}_{2a} |2a\rangle,
\nonumber \\
|\overline{2b}\rangle &=& c_{2a} |2c\rangle - c^{\ast}_{2c} |2a\rangle,
\label{2b}
\end{eqnarray}
where
$c_{2c}$ and $c_{2a}$ are some constants normalized such that
$|c_{2c}|^2+|c_{2a}|^2=1$,
and assume that $1c$ is coupled to $2b$ at the step region.
Then, this signifies that
$\overline{2b}$ is orthogonal to $1c$,
and
$\overline{2b}$ represents a branch that is unaffected by the proximity to $1c$.
Indeed,
$\overline{2b}$ state remains flat, insensitive to the flux, 
and there is no anti-crossing between $1c$ and $\overline{2b}$.
Only at $\phi=\pi$ it reunites with $2b$.
In contrast to the flat $\overline{2b}$-subband,
$1c$ and $2b$  show a clear anti-crossing feature.
The two branches of the anti-crossing may be presented as
\begin{eqnarray}
|1c2b\rangle &=& c_{1c} |1c\rangle + c^{\ast}_{2b} |2b\rangle,
\nonumber \\
|\overline{1c2b}\rangle &=& c_{2b} |1c\rangle - c^{\ast}_{1c} |2b\rangle,
\end{eqnarray}
where
$c_{1c}$ and $c_{2b}$ are some constants that vary as a function of $\phi$:
$c_{1c}=c_{1c} (\phi)$ and $c_{2b}=c_{2b} (\phi)$.
The constants $c_{1c}$ and $c_{2b}$ are also normalized such that
$|c_{1c}|^2+|c_{2b}|^2=1$.
Let us assume that $|1c2b\rangle$
represents the bonding branch 
focused 
in Fig. \ref{step2_zoom} (a) and (b), 
while $|\overline{1c2b}\rangle$ corresponds to the anti-bonding branch.
To illustrate the evolution of $|\overline{1c2b}\rangle$ as a function of $\phi$,
we have plotted in Fig. \ref{wf_step2},
the spatial profile of the corresponding wave function 
at different values of $\phi/\pi$ ($=0.3, 0.45, 0.6$).
One can see that a dominant weight of the wave function is
on the $N_2$-side at $\phi/\pi=0.3$, which is shifted to the
$N_1$-side as $\phi/\pi$ is increased.

\begin{figure}[htbp]
(a)
\includegraphics[width=80mm, bb=0 0 360 355]{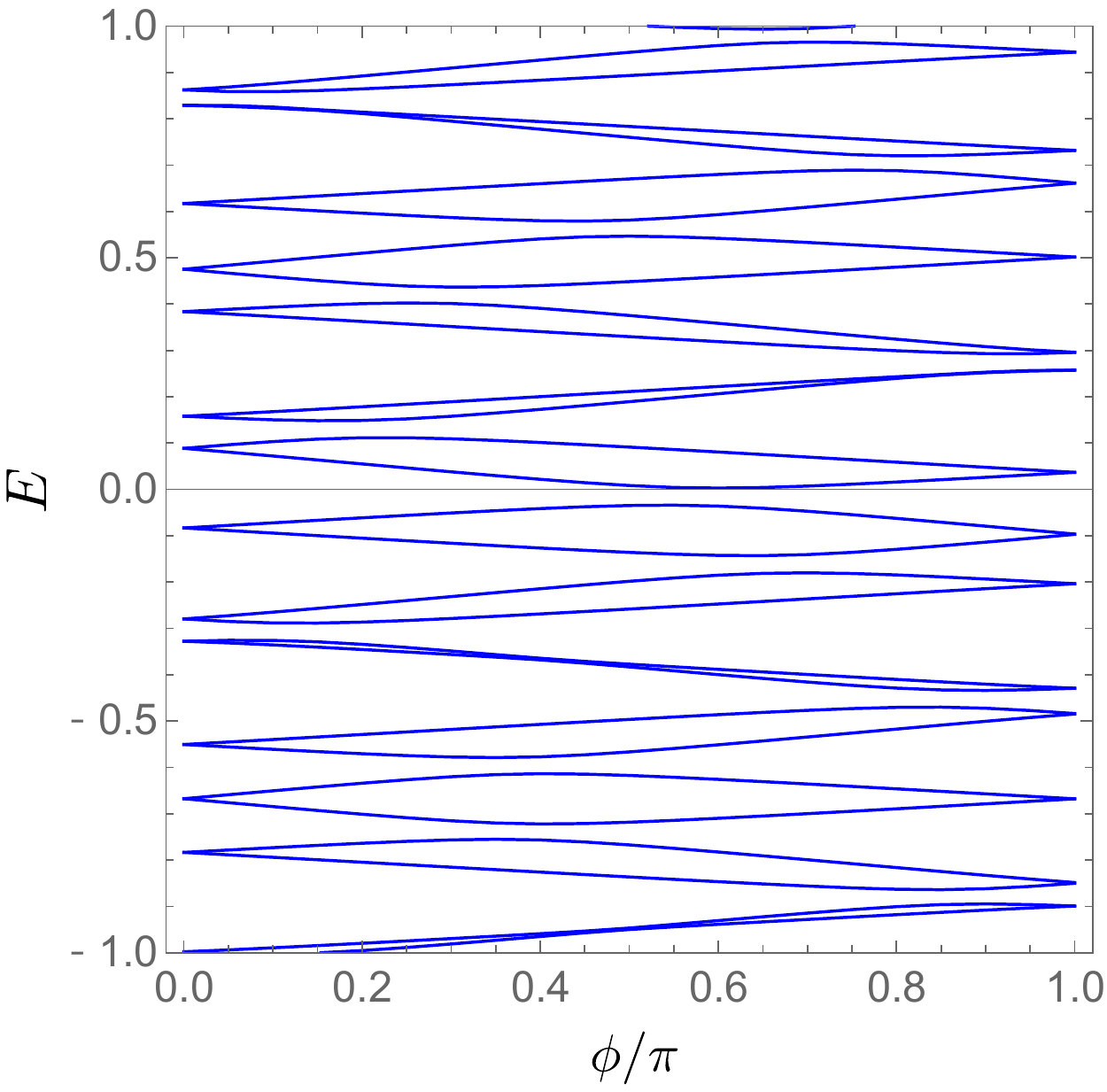}
\\
(b)
\includegraphics[width=80mm, bb=0 0 360 355]{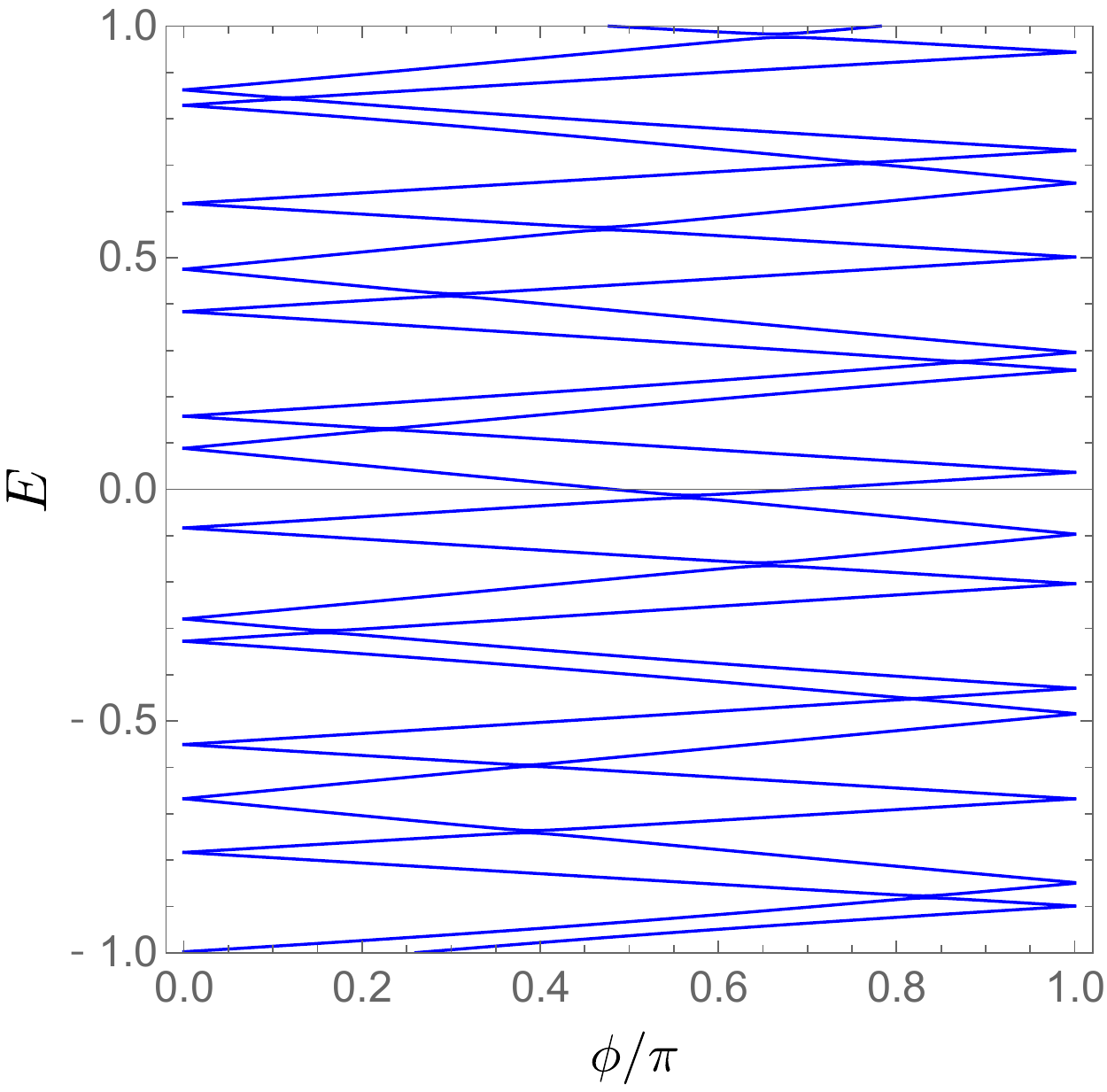}
\vspace{-2mm}
\caption{Spectral flow in the two step geometry 
in the case of the configuration $(\phi_1, \phi_2)$
such that
(a) $(\phi_1, \phi_2) = (\phi, \phi)$.
(b) $(\phi_1, \phi_2) = (\phi, -\phi)$.
}
\label{step2_pm}
\end{figure}

\begin{figure}[htbp]
\includegraphics[width=80mm, bb=0 0 310 281]{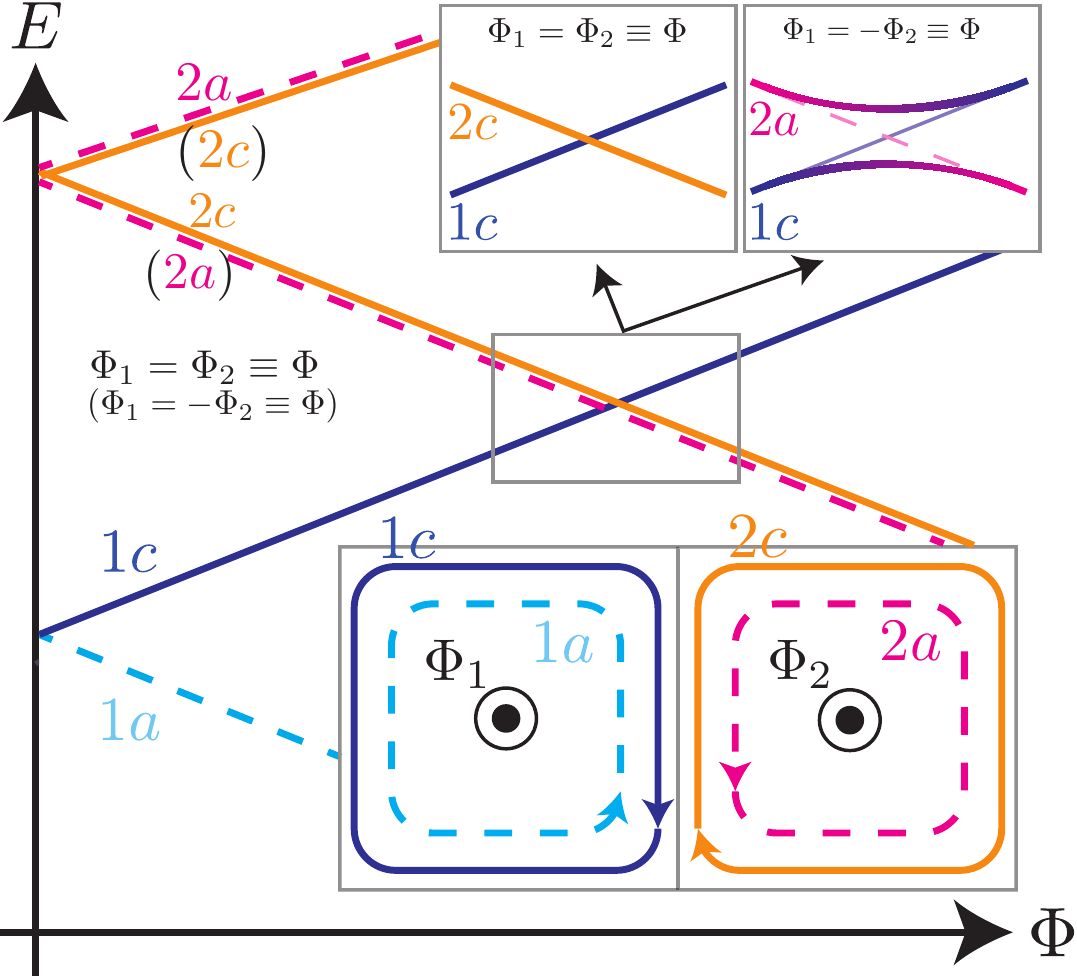}
\vspace{-2mm}
\caption{Illustration of the spectral flow in the two-step geometry
with the flux configuration of (a) $(\phi_1, \phi_2) = (\phi, \phi)$,
and (b) $(\phi_1, \phi_2) = (\phi, -\phi)$,
corresponding, respectively, to panel (a) and (b) in Fig. \ref{step2_pm}.
}
\label{step2_pm_schema}
\end{figure}

In Eq. (\ref{2b}) we assumed simply that $1c$ is coupled to the $2b$-combination,
without actually specifying what a percentage comes from $2c$ in this combination, 
and what a percentage from $2a$.
In this last part,
we clarify this point by studying yet another type of spectral flow.
Two panels of Fig. \ref{step2_pm}
show such a spectral flow
in the cases of flux configurations such that
$(\phi_1, \phi_2)=(\phi, \phi)$ in panel (a),
{\rm i.e.},
case of the flux introduced in the same direction
on the $N_1$ and $N_2$ sides,
while
$(\phi_1, \phi_2)=(\phi, -\phi)$ in panel (b),
{\rm i.e.},
case of the flux inserted in the opposite directions.
Whether the spectrum responds
either in the upward or downward direction 
is a combined effect of
the direction of the propagating 1D channel
and that of the flux introduced.
Therefore, by changing the relative sign of $\phi_2$ with respect to $\phi_1$,
one can bring together different combinations of channels
in the flow of the spectrum
(see Fig. \ref{step2_pm_schema}).

For example, in the $(\phi, \phi)$-configuration,
combinations such as (1c, 2a) and (2c, 1a),
{\rm i.e.}, pairs of co-propagating modes,
get close to one another in the spectral flow.
As one can see in Fig. \ref{step2_pm} (a),
the two 
branches show clear anti-crossing,
making the flow of the spectrum disconnected.
These co-propagating pairs indeed couple 
and recombine at the step region.
On contrary,
in the $(\phi, -\phi)$-configuration [Fig. \ref{step2_pm} (b)],
combinations such as (1c, 2c)
and (1a, 2a),
{\rm i.e.}, a pair of counter-propagating modes,
meet in the spectral flow, 
and show practically no conspicuous anti-crossing.
These imply that
at the step region
the co-propagating combinations such as
(1c, 2a) and (2c, 1a)
are coupled,
while
counter-propagating combinations such as
(1c, 2c) and (1a, 2a)
are not coupled,
{\rm i.e.},
$c_{2c}\simeq 0$, $c_{2a}\simeq 1$
in terms of Eq. (\ref{2b}),
implying actually $|2b\rangle=|2a\rangle$.
The co-propagating combinations
compose the $\phi$-dependent part,
while the counter-propagating combinations 
are responsible for
the $\phi$-independent part,
{\rm i.e.}, the flat part of the
spectral flow shown in Figs. \ref{step2_W2} and \ref{step2_zoom},
ensuring together the connectedness of the nontrivial spectral flow.

In the situation studied so far
we have seen that
the entire circuit is no longer perfectly conducting 
as a result of the coupling between multiple channels.
Yet, there still exists a perfectly conducting channel in the network, 
which is now nontrivially distributed 
in space, extended over coupled circuits (see Fig. 8). 
A prominent feature also manifests in the energy-momentum space 
as a specific type of continuous spectral flow (Fig. 7).

\section{concluding remarks}

We have studied protected helical conducting channels
that appear in WTI nano-flakes,
WTI terraces, 
and WTI nanoarchitectures,
and their robustness against disorder.
After a close inspection of the 
nano-flake case in Sec. III, we have highlighted the cases 
of WTI terraces and steps in Sec. IV.
In contrast to the case of a single step,
an electron incident at the step region
in the two-step geometry
can be transmitted to either side of the step.
To study the nature of the corresponding spectral flow,
encoding the robustness or non-robustness of the helical channel,
was one of the central issues addressed in the paper.
In the geometry studied
an even number of helical channels are incident at the step region,
which can, in principle, be gapped out and get localized in the presence of disorder.
Yet, we find
a nontrivial, connected spectral flow
as a clear signature of the immunity to disorder.
This happens in such a way that each part of the spectrum corresponding to a different 1D channel appear
in segregated energy regions,
but as a whole the spectral flow is connected,
{\rm i.e.},
each piece of the spectrum in a segregated energy region
combines to give the entire connected spectrum.
As a result, 
these even number of channels
are protected and robust against disorder,
alike the case of an odd number of channels.
At a quantum junction 
at each end of the step region 
a non-trivial, energy dependent ``traffic rule'' emerges.

In nano-circuits fabricated on surfaces of a WTI,
a number of 1D channels appear, meet and couple in a nontrivial manner,
forming
nontrivial quantum junctions
at which more than two channels meet in a nontrivial manner.
\cite{JMoore_arxiv}
Here, we have shown an example in which
incident 1D channels stay extended in spite of such coupling.
The obtained results indicate that
robustness of the 1D helical channels
established in simple nano-structures, such as WTI nano-flakes and steps,
is a generic feature
that can be applied to more involved and realistic nanoarchitectures.

\acknowledgments
Y.Y. is supported by JSPS as a Special Doctoral Researcher
and by Grant No. 15J06436.
Y.T. and K.I. are supported by a Grant-in-Aid for Scientific Research 
(B) and (C) 
(Nos. 24540375, 15H0370001, 15K05130 and 15K05131).

\appendix 

\section{Derivation of Eqs. (\ref{Eqk}), (\ref{qm}), (\ref{en_odd})
}

%
The low-enery electron states that appear on the side surface of $1 \le z \le N_{z}$
are described by the following effective Hamiltonian:\cite{arita}
\begin{eqnarray}
   H_{\rm eff} &=& \sum_{z=1}^{N_{z}}
     |z\rangle
       \left[ \begin{array}{cc}
         A_{\parallel}k_{\parallel} & 0 \\
         0 & -A_{\parallel}k_{\parallel}
       \end{array} \right]
     \langle z|
   \nonumber \\
 &&  \hspace{-9mm}
   + \sum_{z=1}^{N_{z}-1}
     \left\{
     |z+1\rangle
       \left[ \begin{array}{cc}
         0 & -\frac{1}{2}A_{z} \\
         \frac{1}{2}A_{z} & 0
       \end{array} \right]
     \langle z|
   + {\rm H.c.}
     \right\} ,
\label{2DDiracHam}
\end{eqnarray}
where $|z\rangle \equiv \left\{|z\rangle_{\uparrow},|z\rangle_{\downarrow}\right\}$
represents two-component state vector for the $z$th 1D helical 
channel, 
and $k_\parallel$ represents a component of the momentum
in the direction the side surface is extended, say, $x$ or $y$.

Eq.(\ref{2DDiracHam}) exhibits two Dirac cones centered at
$(k_{z}, k_{\parallel})=(0, 0)$ and $(\pi, 0)$ in the reciprocal space.
%
%
The wave function for a surface state with $k_\parallel$ 
is expressed as 
\begin{align}
  |\Psi\rangle
  = e^{i k_{\parallel} \zeta}\sum_{z=1}^{N_{z}} 
    |z\rangle
      \left[
        \begin{array}{c}
          \alpha_z \\ \beta_z
        \end{array}
      \right] ,
\label{eigenfunc}
\end{align}
where $\zeta=x$ or $y$. 
In the expression Eq. (\ref{eigenfunc}) 
\begin{align}
   \left[
     \begin{array}{c}
       \alpha_z \\ \beta_z
     \end{array}
   \right]
   = \psi(z)
     \left[ \begin{array}{c}
              a \\
              b
            \end{array}
     \right] ,
\label{Efunc}
\end{align}
where the transverse function $\psi(z)$ must satisfy
the boundary condition of

\begin{equation}
  \psi(0) = 0,\ \ \ 
  \psi(N_{z}+1) = 0 .
\label{bc}
\end{equation}
By superposing plane wave solutions
stemming from 
two Dirac cones: 
$(k_z, k_\parallel)=(q, k_\parallel)$, 
$(\pi-q, k_\parallel)$,
one can construct a wave function 
compatible with Eqs. (\ref{Efunc}) and (\ref{bc}) 
such that 
\begin{eqnarray}
   \left[
     \begin{array}{c}
       \alpha_z \\ \beta_z
     \end{array}
   \right]
  & \propto& (e^{iqz} -e^{i\left(\pi-q\right)z})
   \left[
     \begin{array}{c}
       a \\
       b
     \end{array}
   \right]
   \nonumber \\
   & \propto& (e^{iqz}-(-1)^{z}e^{-iqz})
   \left[
     \begin{array}{c}
       a \\
       b
     \end{array}
   \right],
  \label{psi_z}
\end{eqnarray}
where $^t[a, b]$ satisfies 
\begin{equation}
   \left[
     \begin{array}{cc}
       A_\parallel k_\parallel & i\sin{q} \\
       -i \sin{q} & -A_\parallel k_\parallel
     \end{array}
   \right]
   \left[
     \begin{array}{c}
       a \\ b
     \end{array}
   \right]
=
E
   \left[
     \begin{array}{c}
       a \\b
     \end{array}
   \right] .
\label{EigenEq}
\end{equation}
From Eq. (\ref{EigenEq}), one finds 
\begin{equation}
E=\pm\sqrt{(A_z \sin{q})^2+(A_\parallel k_\parallel)^2}.
\label{Ekk}
\end{equation}

Allowed discrete values of $q$ 
given in Eq. (\ref{qm})
are specified by the second
equality of Eq. (\ref{bc}) 
by the vanishing of $\psi (z)$ at $z=N_z+1$,
{\rm i.e.}
\begin{eqnarray}
  q= {{m}\pi\over N_z+1} \equiv q_{{m}},
\label{q_odd}
\end{eqnarray}
where ${m}$ is either an integer or a half-odd integer.
For an odd $N_z$, 
${m}$ takes an integral value; 
${m}=0, \pm 1, \pm 2, \cdots, \pm {(N_z-1) \over 2}$. 
While for $N_z$ even, 
${m}$ becomes a half-odd integer; 
${m}=\pm{1 \over 2}, \pm{3 \over 2}, \cdots, \pm{{N_z-1}\over 2}$. 
Substituting
Eq. (\ref{q_odd}) into Eq. (\ref{Ekk}), 
one completes
the derivation of
Eqs. (\ref{Eqk}) and (\ref{en_odd}).

\bibliography{mqc_r3v3}

\end{document}